\documentclass[aps,pra,reprint,amsmath,amssymb]{revtex4-2}
\usepackage{bm,graphicx,mathtools,url,tikz}
\usetikzlibrary{decorations.pathmorphing}
\newcounter{explanatory}[section]
\renewcommand{\theexplanatory}{\alph{explanatory}}
\newcommand{\explanatoryheading}[1]{%
  \par\smallskip\refstepcounter{explanatory}%
  \noindent\textit{\theexplanatory.\ #1}\enspace\ignorespaces}
\begin{document}
\title{The Photonic Bath at Infinity: Foundations, Derivations, and Consistency Checks}
\author{Jian-Sheng Wang}
\affiliation{Department of Physics, National University of Singapore, Singapore 117551}
\date{8 November 2026}
\begin{abstract}
The photonic bath at infinity represents the unbounded electromagnetic environment as an effective reservoir in the nonequilibrium Green's function (NEGF) formalism. It accounts for outgoing radiation and incoming electromagnetic fluctuations, allowing radiative transport to be described consistently. We clarify how environmental fluctuations are incorporated without double-counting the outgoing radiation response already contained in the free retarded Green's function. We develop four complementary approaches to deriving the bath self-energy: infinitesimal environmental absorption, semi-infinite leads, inversion of the free Green's function, and sesquilinear forms with outgoing radiation boundary conditions. In particular, we present an explicit three-site matrix inversion for a planar geometry and introduce a sesquilinear form that relates the bath self-energy to the Silver--M\"uller radiation condition. We establish the connections among these approaches and obtain explicit bath self-energies for one-dimensional, planar, and spherical geometries. The resulting bath self-energies reproduce the Planck spectrum, the Stefan--Boltzmann law, and radiation pressure, and provide the radiative contribution required for energy conservation in photon NEGF.
\end{abstract}
\maketitle

\section{Introduction}\label{sec:introduction}
The electromagnetic field transports energy and momentum between material objects and into the surrounding vacuum. Radiative heat transfer is traditionally described by fluctuational electrodynamics, originating from the work of Rytov \cite{Rytov1953}; the classic calculation of Polder and van Hove provides an example \cite{PolderVanHove1971}. In this approach, fluctuating currents generate electromagnetic fields whose energy and momentum fluxes are evaluated using the Poynting vector and the Maxwell stress tensor. An alternative description is provided by nonequilibrium Green's functions (NEGF), in which material response and fluctuations are represented by photon self-energies; see the review by Wang \emph{et al.} \cite{WangPengZhangZhangZhu2023}. The radiation environment must also be included to account for photons escaping to infinity and for thermal radiation incident from the surroundings.

\begin{figure}[t]
\centering

\begin{tikzpicture}[x=1cm,y=1cm,scale=1.10,transform shape,>=stealth, every node/.style={font=\footnotesize}]
  \draw[blue!65!black,thick] (0,0) circle (1.32);
  \filldraw[fill=orange!25,draw=black!65,thick]
    (-.32,-.13) .. controls (-.56,.12) and (-.35,.42) .. (-.09,.37)
    .. controls (.18,.47) and (.45,.25) .. (.33,.02)
    .. controls (.52,-.31) and (.19,-.48) .. (-.10,-.35)
    .. controls (-.31,-.45) and (-.43,-.30) .. cycle;
  \node at (0,-.03) {matter};
  \draw[->,blue!70!black,thick] (0,0) -- (1.03,.82) node[pos=.59,above left] {$R$};
  \node[blue!65!black,anchor=west] at (1.07,.88) {$S_R$};
  \draw[->,blue!65!black,thick,decorate,decoration={snake,amplitude=1.5pt,segment length=7pt}] (1.02,-.83) -- (1.62,-1.32);
  \draw[->,blue!65!black,thick,decorate,decoration={snake,amplitude=1.5pt,segment length=7pt}] (-1.24,.47) -- (-1.78,.68);
  \draw[->,blue!65!black,thick,decorate,decoration={snake,amplitude=1.5pt,segment length=7pt}] (.30,1.29) -- (.43,1.83);
  \node[anchor=west] at (1.42,.05) {exterior vacuum};
  \node[anchor=west] at (1.42,-.23) {$r\to\infty$};
  \node[anchor=west] at (1.40,-.65) {$\Pi_\infty^r$};
\end{tikzpicture}
\caption{Schematic of a material object enclosed by a spherical surface \(S_R\) of radius \(R\). The radiative effect of the exterior electromagnetic field is represented by a bath self-energy associated with the surface. Although the surface is not a physical absorbing shell, the bath accounts for outgoing radiation crossing the surface and supplies incoming thermal radiation through its fluctuations.}
\label{fig:bath-geometry}
\end{figure}
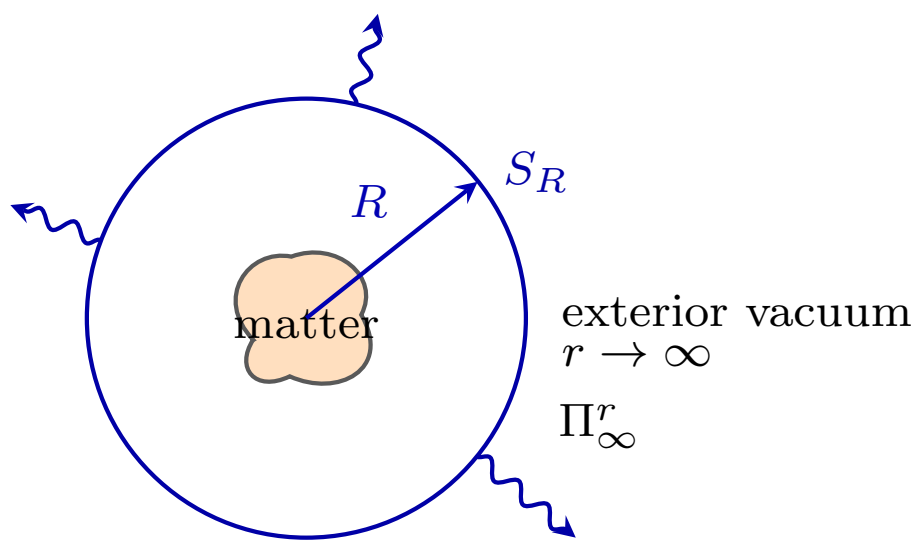

The \emph{photonic bath at infinity} provides such a description. Its role is analogous to that of a semi-infinite lead in electronic or phononic transport: the infinitely extended environment is represented by an effective boundary self-energy. Unlike an ordinary material reservoir, however, the electromagnetic environment is already implicit in the outgoing boundary condition of the free retarded Green's function. Its fluctuations must therefore be represented without counting the outgoing response twice.  Figure~\ref{fig:bath-geometry} illustrates one such geometry: the matter lies inside a closed observation surface, while the surrounding vacuum extends to infinity.   The effect of the eliminated exterior is represented on this surface. 

The general Keldysh equation provides a natural starting point. In addition to material fluctuations, it contains a term describing fluctuations of the free electromagnetic field. For a thermal environment, this term can be rewritten as an effective bath lesser self-energy. The corresponding anti-Hermitian retarded bath operator is
\begin{equation}\label{eq:intro-bath}
\Pi_\infty^r=\frac12\bigl[(v^a)^{-1}-(v^r)^{-1}\bigr],
\end{equation}
where $v^r$ and $v^a$ are the retarded and advanced free photon Green's functions. This expression is straightforward for finite matrices, but its interpretation for differential operators is less obvious. The Maxwell wave operator $L_0$ has a formally Hermitian differential expression; its outgoing and incoming realizations differ through their radiation boundary conditions. Understanding this distinction is an important motivation for the present work.

The treatment of electromagnetic fluctuations in unbounded systems has a long history. Eckhardt emphasized that infinitesimal absorption in an infinite vacuum region can leave a finite contribution if the zero-absorption limit is taken after spatial integration \cite{Eckhardt1984}. Henneberger and collaborators distinguished medium-induced photon fluctuations from those associated with the incident free field \cite{Richter2008,Henneberger2009}. Kr\"uger and collaborators described environmental correlations using scattering operators and the anti-Hermitian part of the inverse free Green's function \cite{Kruger2012}. These works provide essential foundations for the present account.

Within photon NEGF, a spherical bath at infinity was introduced in the treatment of energy, momentum, and angular momentum transport by Zhang \emph{et al.} \cite{ZhangZhuZhangWang2022}, where the NEGF transport formulas were shown to agree with conventional electromagnetic flux calculations. A surface Green's function construction for scalar photons was developed earlier by Peng \emph{et al.} \cite{PengYapZhangWang2017}; its field normalization differs from the transverse vector-potential formulation used here. The planar bath self-energy was used in the Floquet-radiation calculation of Zhu \emph{et al.} \cite{ZhuRenPanTangZhangWang2026}. Tang, Ren, and Wang subsequently gave a systematic derivation of photonic bath operators \cite{TangRenWang2026}.

The purpose of this paper is to establish the connections among different approaches to treating the radiation environment within photon NEGF.
We present four complementary approaches in a unified pedagogical treatment. The first introduces infinitesimal absorption in the exterior and takes the zero-absorption limit after integration. The second eliminates a semi-infinite photon lead using its surface Green's function. The third directly inverts the free Green's function sampled at discrete spatial points. We give particular attention to the full Cartesian three-site matrix inversion, including the longitudinal contact term that is essential for obtaining the correct bath self-energy. The fourth approach uses sesquilinear forms and the Silver--M\"uller radiation condition to explain the anti-Hermitian boundary contribution associated with the Maxwell wave operator.

We obtain results for one-dimensional, planar, and spherical geometries and check them against established electromagnetic results, including the Planck spectrum, the Stefan--Boltzmann law, radiation pressure, and energy conservation. The Meir--Wingreen and Landauer transport formulas provide the corresponding reservoir perspective \cite{MeirWingreen1992,Landauer1957}. We focus on stationary linear electrodynamics and distinguish the purely radiative bath from the complete exterior embedding operator at a finite boundary. The main text develops the physical motivation and intermediate algebra; additional calculations and historical notes appear in the appendices.

\section{Conventions and notation}\label{sec:conventions}
We work in SI units and the temporal (Weyl) gauge, $\phi=0$ \cite{Kay2022}, with
$\mathbf E=-\partial_t\mathbf A$ and $\mathbf B=\nabla\times\mathbf A$.
Our Fourier convention is
\begin{equation}
 \tilde f(\omega)=\int_{-\infty}^{\infty}\!\!dt\,e^{i\omega t}f(t),\quad
 f(t)=\int_{-\infty}^{\infty}\!\!\frac{d\omega}{2\pi}e^{-i\omega t}\tilde f(\omega).
\end{equation}
Greek indices $\mu,\nu$ label Cartesian components; $\alpha,\beta$ label material reservoirs. We write $\hat 1$ for the abstract identity operator, $I$ for the Cartesian identity, and $I_\perp=I-\hat{\mathbf z}\hat{\mathbf z}$ for the tangential identity in planar geometry. When restricted to tangential components, $I_\perp$ is understood as the two-dimensional identity matrix.
The free Maxwell wave operator $L_0$, in the convention of the forthcoming textbook by Wang, Xu, and Zhou \cite{WangXuZhouInPreparation}, is formally written as
\begin{align}
 L_0&=-\epsilon_0\partial_t^2 I-\mu_0^{-1}\nabla\times\nabla\times,\\
 L_0(\omega)&=\epsilon_0\omega^2 I-\mu_0^{-1}\nabla\times\nabla\times.
\end{align}
Strictly speaking, the displayed formula specifies a differential expression; its operator realization also requires a domain, including the appropriate boundary conditions (see Sec.~\ref{sec:forms}). The retarded free Green's function operator $v^r$ satisfies $L_0 v^r=\hat 1$ with the causal outgoing prescription, and the source convention is $L_0\mathbf A=-\mathbf j$. In free space, $v^r=-\mu_0 G^r$, where $G^r$ is the conventional outgoing electromagnetic dyadic Green tensor.
The contour Green's function is defined within the Keldysh formalism \cite{Keldysh1965,HaugJauho2008}:
\begin{equation}
 D_{\mu\nu}(1,2)=\frac{1}{i\hbar}\langle\mathcal T_C\,\delta A_\mu(1)\delta A_\nu(2)\rangle,
\end{equation}
where $\delta\mathbf A=\mathbf A-\langle\mathbf A\rangle$; for zero mean field, $\delta\mathbf A=\mathbf A$. The lesser component is $D^<_{\mu\nu}(1,2)=-(i/\hbar)\langle\delta A_\nu(2)\delta A_\mu(1)\rangle$, and $D^a=(D^r)^\dagger$ in the stationary frequency representation.
The material Dyson equation and general Keldysh equation are
\begin{align}
 D^r&=v^r+v^r\Pi_{\rm mat}^rD^r,\\
 D^<&=D^r\Pi_{\rm mat}^<D^a+(\hat 1+D^r\Pi_{\rm mat}^r)v^<(\hat 1+\Pi_{\rm mat}^aD^a).
\end{align}
The product of the form $D^r \Pi D^a$ means matrix multiplication over Cartesian component indices and convolution in positions at fixed
frequency, i.e.,  
\begin{equation}
\int \!\!d^3{\mathbf r}_1 \int\!\! d^3{\mathbf r}_2 \sum_{\lambda\rho}\!\!D^r_{\mu\lambda}(\omega, {\mathbf r}, {\mathbf r}_1) \Pi_{\lambda\rho}(\omega, {\mathbf r}_1, {\mathbf r}_2)
D^a_{\rho\nu}(\omega, {\mathbf r}_2, {\mathbf r}'). \nonumber
\end{equation} 
For a thermal incoming free field, the equilibrium fluctuation--dissipation relation \cite{CallenWelton1951,Kubo1966} gives $v^<=N_\infty(v^r-v^a)$, with $N_\infty(\omega)=[\exp(\beta_\infty\hbar\omega)-1]^{-1}$ and $N_\infty(-\omega)=-[1+N_\infty(\omega)]$. The equivalent effective lesser self-energy is
\begin{equation}\label{eq:pinf-less}
 \Pi_\infty^<=N_\infty\big[(v^a)^{-1}-(v^r)^{-1}\big].
\end{equation}
We define the anti-Hermitian radiative bath operator by
\begin{equation}\label{eq:conventions-bath}
 \Pi_\infty^r=\frac12\big[(v^a)^{-1}-(v^r)^{-1}\big],\quad
 \Gamma_\infty=i(\Pi_\infty^r-\Pi_\infty^a).
\end{equation}
This operator is used to represent free-field fluctuations; it is \emph{not} added again to the material Dyson equation when $v^r$ already incorporates the outgoing-wave boundary condition. Operator inverses and adjoints require a specified domain, boundary prescription, or consistent finite-dimensional representation.
For positive frequency in a planar vacuum region, define $k_0=\omega/c$, $q_\perp=|\mathbf q_\perp|$, and $k=\sqrt{k_0^2-q_\perp^2}>0$ for propagating modes. For evanescent modes, the outgoing branch has $k=i\kappa$, $\kappa>0$. At negative frequencies, we invoke the symmetry $\Pi^r(-\omega) = \Pi^r(\omega)^*$ and maintain the outgoing or decaying boundary condition at infinity. 
The bath self-energies given below are boundary kernels (with the corresponding surface measure), not three-dimensional bulk kernels. Planar Fourier transforms use $f(\mathbf q_\perp,z)=\int d^2\rho\,e^{-i\mathbf q_\perp\cdot\boldsymbol\rho}f(\boldsymbol\rho,z)$ and inverse measure $d^2q_\perp/(2\pi)^2$.

\section{Photon NEGF and the missing radiation bath}\label{sec:negf}
The electromagnetic environment differs from a conventional material reservoir: even in the absence of matter, the free photon field possesses propagating degrees of freedom and a nonvanishing spectral function. A description that keeps only material polarization fluctuations therefore need not reproduce the incoming thermal radiation. This distinction is particularly important for open systems; a broader introduction to photon NEGF and its transport applications is given in Ref.~\cite{WangPengZhangZhangZhu2023}.

\explanatoryheading{Why a material-only Keldysh equation is insufficient.}
Imagine first that all material objects have been removed.  The free retarded Green's function $v^r$ is certainly not zero: a localized classical source still launches an outgoing electromagnetic wave.  Likewise, a vacuum region in thermal equilibrium supports photons incident from arbitrarily far away.  Yet a lesser Green's function constructed only from material polarization fluctuations, $D^r\Pi_{\rm mat}^<D^a$, would vanish when $\Pi_{\rm mat}^<=0$.  This simple thought experiment reveals the missing ingredient.  The free radiation field has an initial statistical state of its own, independent of the matter fluctuations.

There is a useful analogy with electronic transport.  A lead has two jobs: it supplies the outgoing boundary condition through its retarded embedding self-energy, and it injects particles through its lesser self-energy.  For photons, the free outgoing Green's function already handles the first role, whereas the second remains visible as the additional term in the general Keldysh equation.  Our objective is to put that term into the same reservoir bookkeeping used for the material objects, without modifying the outgoing propagator a second time.

\subsection{Two origins of field fluctuations}
We split the total material polarization into contributions from distinct objects, $\Pi_{\rm mat}=\sum_\alpha\Pi_\alpha$. With the convention of Sec.~\ref{sec:conventions}, the contour Dyson equation gives the retarded equation and the general Keldysh equation quoted there. For clarity, we write the latter as
\begin{equation}\label{eq:Keldysh-full}
D^<=D^r\Pi_{\rm mat}^<D^a+\mathcal W^r v^<\mathcal W^a,
\end{equation}
where $\mathcal W^r=\hat 1+D^r\Pi_{\rm mat}^r$ and $\mathcal W^a=\hat 1+\Pi_{\rm mat}^aD^a$. These electromagnetic dressing operators describe generalized dielectric screening: $\mathcal W^r=(\varepsilon^r)^{-1}$, where $\varepsilon^r=\hat 1-v^r\Pi_{\rm mat}^r$, and $\mathcal W^a=(\mathcal W^r)^\dagger$. The first contribution originates in material fluctuations; the second represents incoming free-field fluctuations dressed by the material response. This separation is closely related to the medium-induced and vacuum-induced contributions discussed by Henneberger and coworkers \cite{Richter2008,Henneberger2009}.
The retarded Dyson equation implies
\begin{equation}
\mathcal W^r=D^r(v^r)^{-1},\qquad
\mathcal W^a=(v^a)^{-1}D^a.
\end{equation}
Consequently, the second term in Eq.~(\ref{eq:Keldysh-full}) can be expressed in reservoir form without changing the retarded propagator,
\begin{align}\label{eq:Keldysh-reservoir}
D^<&=D^r\left(\Pi_{\rm mat}^<+\Pi_\infty^<\right)D^a,\\
\Pi_\infty^<&=(v^r)^{-1}v^<(v^a)^{-1}.
\end{align}
For a thermal free field the fluctuation--dissipation relation yields Eq.~(\ref{eq:pinf-less}) and $\Pi_\infty^<=N_\infty(\Pi_\infty^r-\Pi_\infty^a)$. At positive frequency, the corresponding broadening $\Gamma_\infty$ is positive semidefinite for a passive outgoing radiation channel.

\subsection{What is, and is not, an extra Dyson self-energy}
The free $v^r$ already obeys the outgoing radiation condition. Equation~(\ref{eq:Keldysh-reservoir}) reorganizes the free-field \emph{fluctuation} term; it does not authorize adding $\Pi_\infty^r$ once more to $(D^r)^{-1}=(v^r)^{-1}-\Pi_{\rm mat}^r$. One may instead formulate a finite-domain Dyson problem with an explicit exterior embedding operator, but that is a different partition and must be constructed consistently. Its Hermitian reactive part, particularly in evanescent channels, need not vanish.
The anti-Hermitian construction is therefore best regarded as the spectral radiation operator associated with a specified outgoing free Green's function. In a finite sampled representation it is an ordinary matrix; for differential operators its adjoint also depends on the boundary domain. We specify the representation in the geometrical derivations that follow.

\section{Infinitesimal absorption and the general bath operator}\label{sec:dust}
Before introducing another formal identity, it is useful to state the physical puzzle. An object can emit photons into empty space even though the material polarization vanishes everywhere outside the object. If the exterior is absent from the list of reservoirs, the outgoing energy has no corresponding receiving reservoir in a Meir--Wingreen balance. The retarded vacuum Green's function already describes this outgoing wave, but the corresponding free-field \emph{fluctuations} remain as a separate term in the general Keldysh equation. Our task is to express that term in the same reservoir language used for matter.

We first consider two complementary ways to do so. One is to introduce infinitesimal absorption in the exterior and integrate over its infinite volume. The other is to sample the free Green's function at a finite set of points and examine the anti-Hermitian part of its inverse. We shall show how these procedures lead to a surface contribution and also explain what they do \emph{not} imply about the complete embedding response at a finite boundary.

\explanatoryheading{Why the exterior cannot be replaced by zero absorption too early.}

Consider a finite radiating object surrounded by empty space.  A wave leaving the object may travel arbitrarily far before it is absorbed.  If the absorption rate is $\eta$, the amplitude attenuation length is $c/\eta$.  A local dissipative weight proportional to $\eta$ therefore samples a spatial region whose size diverges as $\eta\to0^+$.  A small local response does not imply a small \emph{integrated} response.  This elementary order-of-limits issue is the physical reason for retaining the exterior dust until the volume integral is evaluated.

The inverse Green's function construction addresses a related but distinct question: after eliminating the exterior, which part of the resulting operator represents an irreversible radiation channel?  For a finite matrix, the difference between an inverse and its adjoint isolates that spectral contribution.  In a continuum description, one must also specify outgoing versus incoming boundary conditions; the formal bulk differential expression alone does not retain this information.

\subsection{Resolvent identity}
Let $L^r=(D^r)^{-1}$ and $L^a=(D^a)^{-1}$ be compatible inverses. Then
\begin{equation}\label{eq:resolvent}
D^r-D^a=D^r(L^a-L^r)D^a.
\end{equation}
Applied to the full photon Green's function in Sec.~\ref{sec:conservation}, this resolvent identity leads to the electromagnetic optical theorem. The identity is elementary for finite matrices. In an open continuum system, however, the domain and the radiation prescription must be retained when interpreting $L^a-L^r$. A differential expression with real coefficients can have distinct retarded and advanced realizations. The corresponding spectral difference is not captured by subtracting the formal bulk differential expressions alone.
For the free field, define
\begin{equation}\label{eq:free-spectral}
\mathcal Q_\infty=(v^a)^{-1}-(v^r)^{-1},\quad
v^r-v^a=v^r\mathcal Q_\infty v^a.
\end{equation}
Our bath operator is $\Pi_\infty^r=\mathcal Q_\infty/2$. It is anti-Hermitian; $\mathcal Q_\infty$ is not the complete exterior embedding operator.
\subsection{The environmental dust construction}
Before assigning a photon self-energy to the environmental dust, let us derive its relation to the conductivity. In the Weyl gauge, the electric field is $\mathbf E=-\partial_t\mathbf A$. An isotropic, local Ohmic response therefore obeys
\begin{equation}
\mathbf j_{\rm ind}=\sigma\mathbf E=-\sigma\partial_t\mathbf A.
\end{equation}
Our Fourier convention is $\mathbf A(t)=\int(d\omega/2\pi)\,e^{-i\omega t}\mathbf A(\omega)$. Hence $\mathbf j_{\rm ind}(\omega)=i\omega\sigma\mathbf A(\omega)$. With the photon self-energy defined by $\mathbf j_{\rm ind}=-\Pi^r\mathbf A$, we find
\begin{equation}\label{eq:dust-conductivity}
\Pi_{\sigma}^r(\omega)=-i\omega\sigma I.
\end{equation}
For real, positive $\sigma$ and $\omega>0$, this has a negative imaginary part, as required for passive absorption. This relation is the starting point of the dust construction; the conductivity is introduced only in the exterior region and is eventually taken to zero. Here $\Pi_\sigma^r$ denotes the coefficient of a local multiplication operator. In a three-dimensional coordinate-space kernel, Eq.~(\ref{eq:dust-conductivity}) also carries $\delta(\mathbf r-\mathbf r' )$.

Why does such a weak conductor attenuate a photon? In a homogeneous region, the wave equation for the transverse vector potential $\mathbf A$ reads
\begin{equation}
\left[\nabla^2+k_\sigma^2\right]\mathbf A=0,\qquad
k_\sigma^2=\frac{\omega^2}{c^2}+i\mu_0\sigma\omega.
\end{equation}
We choose the outgoing root with $\operatorname{Im}k_\sigma>0$. When $\sigma/(\epsilon_0\omega)\ll1$, a first-order expansion gives
\begin{align}
k_\sigma&=\frac{\omega}{c}\sqrt{1+i\frac{\sigma}{\epsilon_0\omega}}\nonumber\\
&\simeq\frac{\omega}{c}+i\frac{\sigma}{2\epsilon_0c}.
\end{align}
Thus an outgoing wave proportional to $e^{ik_\sigma r}$ has the amplitude attenuation factor
\begin{equation}
e^{ik_\sigma r}\simeq e^{i\omega r/c}
\exp\!\left[-\frac{\sigma r}{2\epsilon_0c}\right].
\end{equation}
The field intensity, and therefore the product of retarded and advanced far-field propagators, decays with twice the amplitude attenuation exponent.

It is convenient to parametrize the infinitesimal conductivity by $\sigma=2\epsilon_0\eta$, where $\eta>0$ has dimensions of inverse time. The preceding formulas become
\begin{align}\label{eq:dust-pi}
\Pi_\eta^r&=-2i\epsilon_0\eta\omega,\\
\Pi_\eta^r-\Pi_\eta^a&=-4i\epsilon_0\eta\omega,\\
k_\sigma&\simeq(\omega+i\eta)/c.
\end{align}
Accordingly, the outgoing amplitude decays as $e^{-\eta r/c}$ and the intensity as $e^{-2\eta r/c}$. The amplitude attenuation length is $c/\eta$, which diverges when the conductivity vanishes. Nevertheless, an infinitely extended region can still produce a finite integrated contribution.
For an exterior region $V_{\rm ext}$, the contribution of these weakly absorbing degrees of freedom to a field correlation is
\begin{equation}\label{eq:dust-volume}
\mathcal C_{\rm ext}=\lim_{\eta\to0^+}D_\eta^r\left[\Pi_\eta^r-\Pi_\eta^a\right]_{V_{\rm ext}}D_\eta^a.
\end{equation}
The subscript \(V_{\rm ext}\) indicates an integration over the exterior volume, including the spatial delta function implicit in the local dust self-energy. The limit must be taken \emph{after} that integration: the vanishing absorption can be compensated by the diverging absorption length. This is the point emphasized by Eckhardt in his discussion of unbounded vacuum regions \cite{Eckhardt1984}, and the dust construction is also used in the scattering formulation of Kr\"uger and coworkers \cite{Kruger2012}.
In a simple far-field channel, the radial integral has the form
\begin{equation}\label{eq:dust-radial}
\lim_{\eta\to0^+}\frac{2\eta}{c}\int_0^\infty ds\,e^{-2\eta s/c}=1.
\end{equation}
The exact tensor and geometrical factors are supplied by the propagators. Equation~(\ref{eq:dust-radial}) explains why setting $\eta=0$ before integration would incorrectly eliminate the environmental contribution.

\explanatoryheading{How to read the dust integral.}
In Eq.~(\ref{eq:dust-volume}), the two Green's functions describe propagation from an observation point to a weakly absorbing exterior point and back.  The middle factor measures the spectral absorption at that exterior point.  The product is integrated over every possible exterior position.  It is the same structure as a reservoir correlation $D^r\Pi^<D^a$, except that the reservoir is initially spread over an infinite volume.  Replacing that volume by an effective boundary operator is therefore a reduction of degrees of freedom, not the introduction of an additional physical medium.

The dust method identifies the radiative contribution through an exterior volume integral. To see explicitly how the same contribution can be represented at the boundary, we next turn to a one-dimensional problem, where the outgoing Green's function and its sampled inverse can be calculated directly.

\section{One-dimensional radiation baths}\label{sec:oneD}
The one-dimensional example introduces the central mechanism before we encounter polarization tensors. Imagine observing the field only in a finite interval, while the two exterior half-lines are left unobserved. We seek an operator acting at the two endpoints that reproduces the spectral effect of these eliminated half-lines. The answer should not depend on the arbitrary length of the observed interval.

We first consider a single transverse component $A(z,t)$ of the vector potential. In vacuum, the homogeneous wave equation for the vector potential and the equation for the free retarded Green's function, in our normalization, are
\begin{align}
(\partial_t^2-c^2\partial_z^2)A&=0,\\
\epsilon_0(\omega^2+c^2\partial_z^2)v^r(z,z';\omega)&=\delta(z-z').
\end{align}
At positive frequency, with $k=\omega/c$, the outgoing solution is
\begin{equation}\label{eq:1d-free}
v^r(z,z';\omega)=-\frac{i}{2\epsilon_0c^2k}e^{ik|z-z'|}.
\end{equation}
Indeed, $(\partial_z^2+k^2)e^{ik|z-z'|}=2ik\delta(z-z')$, which fixes both the sign and the coefficient. We take the radiation baths to act at the endpoints of a finite sampled interval; the following construction makes this statement precise.

\explanatoryheading{Why sampling is useful.}
Instead of starting with an inverse differential operator on an unbounded domain, choose a finite set of observation points and evaluate the \emph{known} outgoing Green's function between them.  This produces an ordinary finite matrix.  We can invert it without specifying functional-analytic domains, and we can ask which matrix entries remember that radiation may escape beyond the first and last points.  The result is simple: the anti-Hermitian part of the inverse is confined to those two endpoints.

These points are sampling locations, not atoms in a physical photon lattice.  Their separations need not be equal, and the result below does not require sending the sampling distance to zero.  This distinction is important because the alternative discretized-chain derivation \emph{does} involve a genuine lattice-spacing continuum limit.

\subsection{Inverse of the sampled free Green's function}
Choose $N\geq2$ distinct ordered sites $z_1<\cdots<z_N$ and set $d_j=z_{j+1}-z_j$. The sampled Green's function matrix factorizes as $v^r=gM$, where
\begin{equation}
g=-\frac{i}{2\epsilon_0c^2k},\qquad M_{ij}=e^{ik|z_i-z_j|}.
\end{equation}
For generic spacings $\sin(kd_j)\ne0$, the inverse $M^{-1}$ is tridiagonal. For example, for three sites $z_1=-a,z_2=0,z_3=b$, writing $u=e^{ika}$ and $w=e^{ikb}$,
\begin{widetext}
\begin{equation}\label{eq:threeM}
M=\begin{pmatrix}1&u&uw\\u&1&w\\uw&w&1\end{pmatrix},\quad
M^{-1}=\begin{pmatrix}
\frac1{1-u^2}&-\frac{u}{1-u^2}&0\\
-\frac{u}{1-u^2}&\frac1{1-u^2}+\frac1{1-w^2}-1&-\frac{w}{1-w^2}\\
0&-\frac{w}{1-w^2}&\frac1{1-w^2}
\end{pmatrix}.
\end{equation}
\end{widetext}
Using $1/(1-e^{2ix})=\frac12+\frac{i}{2}\cot x$ gives
\begin{widetext}
\begin{equation}\label{eq:oneD-inverse}
(v^r)^{-1}=\epsilon_0c^2k\begin{pmatrix}
-\cot(ka)+i&\csc(ka)&0\\
\csc(ka)&-\cot(ka)-\cot(kb)&\csc(kb)\\
0&\csc(kb)&-\cot(kb)+i
\end{pmatrix}.
\end{equation}
\end{widetext}
Only the endpoint diagonal entries have imaginary parts. Consequently, the radiative contribution, as defined in Eq.~(\ref{eq:conventions-bath}), is
\begin{equation}\label{eq:1d-bath}
\Pi_\infty^r=-i\epsilon_0c^2k\bigl(|1\rangle\langle1|+|N\rangle\langle N|\bigr).
\end{equation}
The bath self-energy \emph{per endpoint} is therefore $-i\epsilon_0c\omega$. For arbitrary $N$, the same result follows from the tridiagonal inverse: the interior diagonal terms and all off-diagonal terms are real. 

\explanatoryheading{A second route to the same endpoint coefficient.}
The matrix inverse is economical, but it does not show where the incoming thermal fluctuations originate.  The dust calculation supplies that physical interpretation.  One imagines an exterior half-line filled with arbitrarily weak absorbers at the radiation-bath temperature.  The lesser polarization of each absorber is small, but the cumulative field arriving at the boundary remains finite.  Showing that the two approaches agree is valuable because one method emphasizes algebraic elimination while the other emphasizes fluctuation sources.

\subsection{One-dimensional environmental dust}
Before evaluating the dust integral, we explain why the free Green's function suffices even when matter is present. Let $v_\eta^{r,a}$ denote the free propagators with infinitesimal absorption in the exterior, and let $D_\eta^{r,a}$ include the material polarization. The Dyson equation gives
\begin{equation}
D_\eta^r=\mathcal W_\eta^r v_\eta^r,\qquad
D_\eta^a=v_\eta^a\mathcal W_\eta^a,
\end{equation}
where $\mathcal W_\eta^r=\hat 1+D_\eta^r\Pi_{\rm mat}^r$ and $\mathcal W_\eta^a=\hat 1+\Pi_{\rm mat}^aD_\eta^a$. Write $\Delta\Pi_{\eta,\rm ext}= (\Pi_\eta^r-\Pi_\eta^a)_{\rm ext}$. The exterior contribution then factorizes as
\begin{align}
&D_\eta^r\Delta\Pi_{\eta,\rm ext}D_\eta^a\notag\\
&\quad=\mathcal W_\eta^r
\bigl(v_\eta^r\Delta\Pi_{\eta,\rm ext}v_\eta^a\bigr)
\mathcal W_\eta^a.
\end{align}
The bracketed product contains the exterior integral, which can therefore be evaluated using free propagators. Under the limiting-absorption prescription, the bracketed expression becomes $v^r\mathcal Q_\infty v^a$, so the dressed contribution is $D^r\mathcal Q_\infty D^a$. The bath operator is thus independent of the material response, while the resulting field correlation depends on the material response. This factorization is also used in Ref.~\cite{TangRenWang2026}.

As an independent check, introduce an infinitesimal conductivity $\sigma=2\epsilon_0\eta$ in the exterior right half-line. At positive frequency, the outgoing wave number becomes $k_\eta=k+i\eta/c+O(\eta^2)$. At the endpoint, the exterior contribution to the free spectral function is
\begin{align}
\mathcal C_R&=\lim_{\eta\to0^+}\int_0^\infty ds\,
|g|^2 e^{-2\eta s/c}(-4i\epsilon_0\eta\omega)\nonumber\\
&=-\frac{i}{2\epsilon_0c\omega}.
\end{align}
Here $|g|^2=(4\epsilon_0^2c^2\omega^2)^{-1}$ and $\int_0^\infty ds\,e^{-2\eta s/c}=c/(2\eta)$. On the other hand, a local boundary spectral operator $\mathcal Q_R=-2i\epsilon_0c\omega$ gives $g\mathcal Q_Rg^*=-i/(2\epsilon_0c\omega)$, precisely the same result. Thus the one-dimensional bath self-energy is $\mathcal Q_R/2=-i\epsilon_0c\omega$. The left half-line gives an identical endpoint operator. This is a direct illustration of the noncommuting infinite-volume and zero-absorption limits.

The same endpoint coefficient can also be obtained directly from the outgoing radiation boundary condition. We present this alternative derivation using sesquilinear forms in Sec.~\ref{sec:forms}.

\subsection{Discretized semi-infinite chain and earlier work}
The centered finite-difference approximation to the wave operator is
\begin{equation}
\epsilon_0\left[\omega^2 A_j+\frac{c^2}{h^2}(A_{j+1}-2A_j+A_{j-1})\right]=0.
\end{equation}
The retarded outgoing lattice root $\lambda=e^{i\vartheta}$ obeys $2\cos\vartheta=2-(\omega h/c)^2$ and $\vartheta=\omega h/c+O(h^3)$. Eliminating a semi-infinite chain gives a boundary response proportional to $\lambda$. Its large real part cancels the boundary contribution of the discrete central wave operator. 
The remaining term gives the bath self-energy \(-i\epsilon_0c^2k\) in the continuum limit, after accounting for the lattice-spacing factor and the endpoint diagonal.
Appendix~\ref{app:surface} gives the complete discrete bookkeeping; in particular, the embedding self-energy alone should not be confused with the renormalized continuum boundary operator.

\explanatoryheading{Why the surface method needs a boundary correction.}
The finite-difference Laplacian contains two nearest-neighbor couplings at an interior site, but an endpoint of the retained region has only one neighbor inside it.  When the missing half-line is eliminated, the embedding self-energy must be combined with this changed diagonal term.  Taking the continuum limit of the embedding term in isolation leaves a large real contribution.  The physical radiation boundary condition emerges only from the consistently assembled boundary equation.  The cancellation is shown explicitly in Appendix~\ref{app:surface}; the key lesson is to compare complete boundary equations rather than isolated self-energy terms.

\explanatoryheading{Relation to the earlier scalar-photon construction.}
Peng \emph{et al.} \cite{PengYapZhangWang2017} used a related semi-infinite-chain construction for a scalar photon field. Its Green's function sign and continuum normalization differ from the transverse vector-potential formulation used here. The detailed comparison, including the boundary diagonal correction, is given in Appendix~\ref{app:peng-comparison}.

\section{Planar geometry: propagating channels}\label{sec:planar}
A planar geometry is not merely the scalar one-dimensional problem repeated for three Cartesian components. For oblique incidence, the transverse electromagnetic field has both TE and TM polarizations, and the TM electric field has a normal component. The question is whether the scalar endpoint localization survives and, if so, how the different polarization weights arise without inserting them by hand.

With translational invariance in the $xy$ plane, we write $\mathbf q_\perp=(q_x,q_y)^T$ and $k_0=\omega/c$. The outgoing longitudinal wave number satisfies $k^2=k_0^2-q_\perp^2$, with $k>0$ for positive-frequency propagating waves and $k=i\kappa$ ($\kappa>0$) for evanescent waves. The tangential block of the free photon Green tensor is
\begin{equation}\label{eq:planar-tangent}
v^r_{\parallel}(\mathbf q_\perp;z,z')=\frac{e^{ik|z-z'|}}{2i\epsilon_0c^2k}P,
\qquad P=I_\perp-\frac{\mathbf q_\perp\mathbf q_\perp^T}{k_0^2}.
\end{equation}
For $k\ne0$, the Sherman--Morrison formula yields
\begin{equation}\label{eq:planar-Pinv}
P^{-1}=I_\perp+\frac{\mathbf q_\perp\mathbf q_\perp^T}{k^2}.
\end{equation}
At $N$ sampled planes, the tangential matrix is a Kronecker product, $v^r_\parallel=gM\otimes P$. Hence
\begin{equation}
(v^r_\parallel)^{-1}=g^{-1}M^{-1}\otimes P^{-1}.
\end{equation}
The scalar endpoint result from Sec.~\ref{sec:oneD} now immediately implies, at each end (see also the planar bath used by Zhu \emph{et al.}~\cite{ZhuRenPanTangZhangWang2026}),
\begin{equation}\label{eq:planar-result}
\boxed{\Pi_{\infty,\parallel}^r=-i\epsilon_0c^2k\left(I_\perp+\frac{\mathbf q_\perp\mathbf q_\perp^T}{k^2}\right)}.
\end{equation}
The eigenvalue in the in-plane direction perpendicular to $\mathbf q_\perp$ (TE) is $-i\epsilon_0c^2k$;  in the direction parallel to 
$\mathbf q_\perp$ (TM), it is $-i\epsilon_0c^2k_0^2/k$.  We can write the results as  $\Pi_{\infty,\sigma}^r=-i\omega Y_\sigma$,
$\sigma=\mathrm{TE},\mathrm{TM}$, where $Y_\sigma$ is the electromagnetic wave admittance. Thus both polarizations are obtained from the
tangential Green's-function inverse without assuming separate bath formulas. For $k=i\kappa$, the tangential free Green's function matrix is real symmetric, so its inverse has no anti-Hermitian part. Accordingly, the radiative operator is supported on $k^2>0$; the grazing limit $k\to0$ is singular in the TM eigenvalue and is understood through the appropriate limiting/integrated prescription.

\explanatoryheading{What changes between one and three dimensions?}
Translational invariance in the $(x,y)$ plane allows us to label each electromagnetic channel by its conserved transverse wave vector $\mathbf q_\perp$.  Once that label is fixed, propagation along $z$ is effectively one-dimensional, with longitudinal wave number $k=\sqrt{k_0^2-q_\perp^2}$.  This reduction makes the one-dimensional bath calculation immediately relevant, but the field now carries a vector polarization structure.  In particular, the tangential electric field of a TM wave is not normalized in the same way as that of a TE wave.

For $q_\perp<k_0$, $k$ is real and the channel transports radiation arbitrarily far along $z$.  For $q_\perp>k_0$, $k$ is imaginary and the field decays exponentially.  This physical distinction explains the propagating-mode restriction in the dissipative bath self-energy; it should not be confused with the full response of a finite boundary, which can contain real evanescent terms.

\subsection{Direct planar dust integral}
The same coefficient follows without a sampled matrix. Consider the right exterior half-space, with a weak conductivity $2\epsilon_0\eta$. For $k>0$, the outgoing longitudinal wave number has the first-order expansion
\begin{equation}
k_\eta=k+i\frac{\eta\omega}{c^2k}+O(\eta^2).
\end{equation}
Let $V_+$ be the full three-dimensional transverse projector for the outgoing wave vector $(\mathbf q_\perp,k)$. Since $V_+^2=V_+$, the product of the outgoing propagator and its adjoint contains $V_+$, whose tangential block is $P$. The exterior volume integration at fixed transverse wave vector includes
\begin{equation}
\int_0^\infty dz\,e^{-2\operatorname{Im}k_\eta z}
=\frac{c^2k}{2\eta\omega}+O(1).
\end{equation}
Inserting the dust spectral factor $-4i\epsilon_0\eta\omega$ and $|g|^2=(4\epsilon_0^2c^4k^2)^{-1}$ gives the tangential environmental spectral correlation for one side,
\begin{equation}\label{eq:planar-dust-correlation}
\mathcal C_{R,\parallel}=-\frac{i}{2\epsilon_0c^2k}P.
\end{equation}
This can be reproduced by a local boundary operator $\mathcal Q_R=-2i\epsilon_0c^2kP^{-1}$, since
\begin{equation}
(gP)\,\mathcal Q_R\,(gP)^\dagger
=-\frac{i}{2\epsilon_0c^2k}P.
\end{equation}
Thus $\Pi_{\infty,R}^r=\mathcal Q_R/2$ agrees with Eq.~(\ref{eq:planar-result}). The left exterior gives the same tangential result. This calculation shows explicitly where the reciprocal polarization factor $P^{-1}$ comes from: it is required to reconstruct the spectral correlation after propagating from the boundary.
For $q_\perp>k_0$, $k=i\kappa$ and no radiation propagates to arbitrarily large $|z|$; the anti-Hermitian radiative bath vanishes in this limit. A finite-distance exact embedding response may still contain nonzero real evanescent contributions.

The polarization calculation establishes the TE and TM coefficients, but it uses the tangential block of the Green's function from the outset. The next section starts instead from the full Cartesian Green's function and shows how the same tangential bath emerges when its longitudinal contact term is retained during matrix inversion.

\section{Cartesian matrix inversion and the radiation bath}\label{sec:cartesian}
\subsection{The question behind the calculation}
Suppose that we know the \emph{free} retarded photon Green's function everywhere in empty space. Can we identify, using this Green's function alone, which degrees of freedom behave as reservoirs carrying radiation away from a finite region? This is a useful question because a free retarded Green's function already knows about outgoing radiation, even though it contains no material self-energy. The general Keldysh equation shows that the missing free-field fluctuations can be represented by the anti-Hermitian part of the inverse free Green's function [Eq.~(\ref{eq:conventions-bath})]. We now ask where that anti-Hermitian part resides after we sample the field at a finite number of planes. Throughout this section, the inverse refers to the finite sampled Green's-function matrix, rather than to the inverse Maxwell differential operator discussed in Sec.~\ref{sec:forms}.

A surprising result is that the inverse matrix has imaginary contributions \emph{only at its outermost planes}. These contributions are precisely the two baths at infinity. The word ``infinity'' refers to the eliminated exterior, not to the position of the sampled boundary: moving the two outer sampling planes does not change their radiative self-energy.

Section~\ref{sec:oneD} established the scalar sampled inverse and its endpoint localization. We now use that result to study the full $3\times3$ Cartesian Green's tensor, including the direction-dependent matrices $V_-$, $V_0$, and $V_+$, and the longitudinal contact term. The new issue is the polarization structure, not the scalar spatial inversion.

\subsection{Scalar three-site result}
For the three planes $z_1=-a$, $z_2=0$, and $z_3=b$, Sec.~\ref{sec:oneD} gives the sampled scalar matrix $M$ and its inverse in Eq.~(\ref{eq:threeM}), and the resulting endpoint bath operator in Eq.~(\ref{eq:1d-bath}). In particular, the anti-Hermitian part of the inverse is confined to the two outer planes, with coefficient $-i\epsilon_0c^2k$ at each endpoint. We will use this established result without repeating the $3\times3$ scalar inversion.

\subsection{Why the Cartesian propagator needs three matrices}
For a planar system, we Fourier transform in $x$ and $y$. Choose the transverse wave vector along $y$ for the calculation,
\begin{equation}
\mathbf q_\perp=(0,q)^T,\qquad k_0=\omega/c,\qquad k^2=k_0^2-q^2.
\end{equation}
The outgoing wave vector between two distinct planes is $\mathbf K_\pm=(0,q,\pm k)^T$. The transverse projector $I-\mathbf K_\pm\mathbf K_\pm^T/k_0^2$ consequently has two forms:
\begin{equation}\label{eq:Vpm}
V_\pm=\begin{pmatrix}
1&0&0\\
0&k^2/k_0^2&\mp qk/k_0^2\\
0&\mp qk/k_0^2&q^2/k_0^2
\end{pmatrix}.
\end{equation}
The $yz$ elements reverse sign when the propagation direction is reversed; the $xx$, $yy$, and $zz$ entries do not. For $z\ne z'$, the free Cartesian Green's function is
\begin{equation}\label{eq:inv-ped-dyadic}
v^r(z,z')=g e^{ik|z-z'|}V_{\operatorname{sgn}(z-z')},
\qquad g=\frac{1}{2i\epsilon_0c^2k}.
\end{equation}
Here the first argument is the observation plane and the second the source plane. Since $z_1<z_2<z_3$, the upper-right blocks use $V_-$ and the lower-left blocks use $V_+$. This ordering is fixed by the Green's function; it is not a choice of signs made during inversion.

At coincident planes the regular part is the symmetric limit, in which the mixed $yz$ entries vanish:
\begin{equation}\label{eq:inv-ped-V0}
V_0=\operatorname{diag}(1,k^2/k_0^2,q^2/k_0^2).
\end{equation}
However, $gV_0$ is \emph{not} the complete coincident Green's tensor. The longitudinal $zz$ entry contains a local distributional term, $+\delta(z-z')/(\epsilon_0\omega^2)$ in our convention. Omitting it before matrix inversion is not a harmless simplification: it changes the inverse of the full Cartesian problem.

\subsection{A concrete prescription for the contact term}
To give the sampled contact term a meaning, introduce a positive regularization length $\delta$ and set $\delta(z_i-z_j)\mapsto\delta_{ij}/\delta$. At each sampling plane this adds
\begin{equation}\label{eq:inv-ped-contact}
C_\delta=\frac{1}{\epsilon_0\omega^2\delta}
\begin{pmatrix}0&0&0\\0&0&0\\0&0&1\end{pmatrix}.
\end{equation}
We emphasize the order of operations: $a$, $b$, $\omega$, and $q$ remain fixed; we invert the finite $9\times9$ matrix; only then do we let $\delta\to0^+$. A concrete motivation is to approximate the one-dimensional
Dirac delta by a normalized narrow cell profile,
\begin{equation}\label{eq:inv-cell-delta}
\delta_\delta(z)=\frac{1}{\delta}
\begin{cases}
1,& |z|<\delta/2,\\
0,& \text{otherwise},
\end{cases}
\qquad \int dz\,\delta_\delta(z)=1.
\end{equation}
For distinct sampling planes with nonoverlapping cells,
its diagonal value is $1/\delta$ and its off-diagonal
values vanish, motivating the replacement above.
This remains a \emph{point-sampling regularization}: a Galerkin
cell average with normalized basis functions carries its own
quadrature weights and is not automatically identical to the
sampled matrix.  The limit established below is conditional on
this prescription and on fixed nonsingular sampling positions.

With $z_1=-a$, $z_2=0$, and $z_3=b$, the matrix is
\begin{widetext}
\begin{equation}\label{eq:full9}
v^r=\begin{pmatrix}
gV_0+C_\delta&ge^{ika}V_-&ge^{ik(a+b)}V_-\\
ge^{ika}V_+&gV_0+C_\delta&ge^{ikb}V_-\\
ge^{ik(a+b)}V_+&ge^{ikb}V_+&gV_0+C_\delta
\end{pmatrix}.
\end{equation}
\end{widetext}
This is the full Cartesian version of the three-site construction. Each entry displayed here is itself a $3\times3$ block. The calculation would be cumbersome if we expanded all 81 scalar entries, but its structure allows a short exact inversion.

\subsection{Schur complement: why the normal components disappear}
Reorder the nine variables so that the six tangential components $(A_x,A_y)$ at the three planes come first and the three normal components $A_z$ come last. The matrix becomes
\begin{equation}\label{eq:inv-ped-block}
v^r=\begin{pmatrix}A&B\\C&E_\delta\end{pmatrix},\qquad
E_\delta=\frac{I_3}{\epsilon_0\omega^2\delta}+E_0.
\end{equation}
The finite matrices $B$, $C$, and $E_0$ contain the direction-dependent $yz$ couplings from $V_\pm$. Since the contact term dominates the normal block,
\begin{equation}\label{eq:inv-ped-Einv}
E_\delta^{-1}=\epsilon_0\omega^2\delta I_3+O(\delta^2).
\end{equation}
The exact Schur complement of $E_\delta$ is
\begin{equation}
S_\delta=A-BE_\delta^{-1}C.
\end{equation}
Thus $S_\delta=A+O(\delta)$, and the full inverse is
\begin{widetext}
\begin{equation}\label{eq:inv-ped-schur}
(v^r)^{-1}=\begin{pmatrix}
S_\delta^{-1}&-S_\delta^{-1}BE_\delta^{-1}\\
-E_\delta^{-1}CS_\delta^{-1}&E_\delta^{-1}+E_\delta^{-1}CS_\delta^{-1}BE_\delta^{-1}
\end{pmatrix}.
\end{equation}
\end{widetext}
Provided $A$ is invertible, taking the limit gives
\begin{equation}\label{eq:schur-limit}
\lim_{\delta\to0^+}(v^r)^{-1}
=\begin{pmatrix}A^{-1}&0\\0&0\end{pmatrix}.
\end{equation}
The vanishing normal row and column have now been \emph{derived}, rather than imposed. They arise because the large contact term suppresses the normal block of the inverse. Notice that the sign differences between $V_+$ and $V_-$ were retained throughout; their effects enter $B$ and $C$, but are multiplied by $E_\delta^{-1}=O(\delta)$ and do not survive the stated limit. This does \emph{not} mean that TM radiation lacks a normal electric field. That component remains present in the propagating Green's function and in Maxwell-stress calculations.

The longitudinal contact term thus has a clear physical consequence: in the limit $\delta\to0^+$, the inverse Green's function has no independent normal ($z$) boundary component, and the bath self-energy acts only in the tangential $(x,y)$ plane. This does not exclude a nonzero normal field for TM-polarized radiation; rather, that field is fixed by the Maxwell constraints and the tangential radiative amplitudes. The spherical far-field bath has the analogous tangential structure, expressed by the projector $I-\hat{\mathbf R}\hat{\mathbf R}^T$, which annihilates the local radial direction. In that geometry, transversality follows directly from the asymptotic outgoing radiation field, without an explicit contact-term calculation.

\subsection{The Kronecker product and the TE/TM factors}
What remains is the tangential $6\times6$ matrix $A$. The tangential blocks of $V_+$, $V_-$, and $V_0$ are identical:
\begin{equation}
P=\begin{pmatrix}1&0\\0&k^2/k_0^2\end{pmatrix}.
\end{equation}
All dependence on the positions is therefore carried by the scalar matrix $M$ in Eq.~(\ref{eq:threeM}). This factorization can be written compactly as a Kronecker product,
\begin{equation}\label{eq:inv-ped-kron}
A=gM\otimes P
=g\begin{pmatrix}P&uP&uwP\\uP&P&wP\\uwP&wP&P\end{pmatrix}.
\end{equation}
The notation $\otimes$ simply means that each scalar entry $M_{ij}$ is replaced by the $2\times2$ block $M_{ij}P$. It follows that
\begin{equation}\label{eq:inv-ped-kroninv}
A^{-1}=g^{-1}M^{-1}\otimes P^{-1},\qquad
P^{-1}=\begin{pmatrix}1&0\\0&k_0^2/k^2\end{pmatrix}.
\end{equation}
The first eigenvector of $P$ is along $x$, perpendicular to $\mathbf q_\perp$, and corresponds to TE (s) polarization. The second is along $y$, parallel to $\mathbf q_\perp$, and corresponds to the tangential part of TM (p) polarization. The factor $k_0^2/k^2$ is not guessed from polarization arguments: it comes directly from the inverse of $P$.

Using the endpoint result in Eq.~(\ref{eq:1d-bath}), the bath self-energy at \emph{each} outer plane is
\begin{equation}\label{eq:cart-bath}
\Pi_{\infty,L}^r=\Pi_{\infty,R}^r
=-i\epsilon_0c^2k\begin{pmatrix}
1&0&0\\0&k_0^2/k^2&0\\0&0&0
\end{pmatrix}.
\end{equation}
The middle plane has no anti-Hermitian contribution. Finally, rotate the in-plane axes to arbitrary $\mathbf q_\perp=(q_x,q_y)^T$. The tangential matrix is now $P=I_\perp-\mathbf q_\perp\mathbf q_\perp^T/k_0^2$. Its inverse, by direct multiplication or the Sherman--Morrison identity, is $P^{-1}=I_\perp+\mathbf q_\perp\mathbf q_\perp^T/k^2$. We obtain
\begin{equation}\label{eq:cart-final}
\boxed{\Pi_\infty^r=-i\epsilon_0c^2k
\left(I_\perp+\frac{\mathbf q_\perp\mathbf q_\perp^T}{k^2}\right)\theta(k^2).}
\end{equation}
This is the planar radiative bath formula.  It agrees with
the independently obtained TE and TM endpoint coefficients in
Sec.~\ref{sec:planar}; in the basis parallel and perpendicular to
$\mathbf q_\perp$, its eigenvalues are
$-i\epsilon_0c^2k_0^2/k$ (TM) and
$-i\epsilon_0c^2k$ (TE), respectively.  The independent
polarization calculation therefore checks the coefficient and
tangential structure produced by the contact-term limit. The step function $\theta$ reflects the fact that evanescent modes, with $k=i\kappa$, do not carry energy to arbitrarily distant planes: the sampled free Green's function is then Hermitian and its anti-Hermitian inverse vanishes. At $k=0$, the TM factor is singular and requires a limiting prescription under the transverse-wave-vector integral.

\subsection{What has actually been established?}
The calculation explains four facts simultaneously: (i) why only the two outer sampling planes act as radiation baths; (ii) why the bath coefficient does not depend on the distances $a$ and $b$; (iii) why the self-energy acts on tangential boundary variables; and (iv) why the TM coefficient differs from the TE coefficient. The endpoint property is already visible in the scalar inverse, while the polarization factor is generated by the Kronecker product.

The derivation is exact for nonsingular finite sampled matrices followed by the specified $\delta\to0^+$ limit. It is not a theorem about arbitrary definitions of the inverse continuum Maxwell operator. For the purposes of photon NEGF, however, it supplies a concrete and independently checkable construction of the dissipative bath at infinity. 

\section{Sesquilinear forms and the radiation boundary}\label{sec:forms}
The inverse-matrix method of Sec.~\ref{sec:cartesian} makes the
radiative contribution visible in a finite-dimensional setting.  A
natural question remains: how can the same idea be understood when
$(v^r)^{-1}$ is regarded as a differential operator?  At real frequency, the Maxwell wave operator has a formally Hermitian differential expression. Nevertheless, its outgoing and incoming Green's functions differ because the corresponding operator domains impose different radiation boundary conditions.  Sesquilinear forms provide a convenient
way to retain this information and to identify the radiative term.
For the theory of forms and their associated operators, see Kato~\cite{Kato1976}.

\subsection{From an operator to a form}
Let $\mathcal H$ be a complex Hilbert space, with inner product
$\langle f,g\rangle$ antilinear in its first argument.  A densely
defined linear operator $L:D(L)\subset\mathcal H\to\mathcal H$
determines a sesquilinear form
\begin{equation}\label{eq:form-operator}
 a_L(f,g)=\langle f,Lg\rangle,
 \qquad f\in\mathcal H,\quad g\in D(L).
\end{equation}
The notation $D(L)$ is essential: a differential expression specifies how $L$ acts, while the domain specifies the admissible functions, their regularity, and their boundary conditions. Two operators can therefore have the same differential expression but different domains and different adjoints. This distinction is particularly important for outgoing and incoming electromagnetic waves.
The form is antilinear in $f$ and linear in $g$.  For a finite matrix,
$a_L(f,g)=f^\dagger Lg$; thus no new concept is needed in finite
dimensions.  The adjoint \emph{form} is defined by
\begin{equation}\label{eq:form-adjoint}
 a^\dagger(f,g)=a(g,f)^*,
\end{equation}
with the two argument domains interchanged.  If
$L^\dagger$ exists on the relevant vectors, this agrees with the
usual Hilbert-space adjoint relation
$\langle f,Lg\rangle=\langle L^\dagger f,g\rangle$.
There are not two incompatible definitions of Hermitian conjugation;
the form notation makes it possible, through a suitable weak formulation, to work on a domain larger than
the domain of a second-order differential operator.

\explanatoryheading{Why the form domain can be larger.}
Consider the elementary example $\mathcal H=L^2(0,\ell)$ and
\begin{equation}\label{eq:form-h1}
 a(f,g)=\int_0^\ell f'(z)^*g'(z)\,dz,
 \qquad f,g\in H^1(0,\ell).
\end{equation}
Here $H^1$ consists of square-integrable functions whose first weak derivatives are square-integrable; for Sobolev spaces and boundary-value problems, see Lions and Magenes~\cite{LionsMagenes1972}.  The integral is well
defined even if $g''$ does not exist as an $L^2$ function.  For
sufficiently smooth functions, integration by parts gives
\begin{equation}\label{eq:form-ibp}
 a(f,g)=-\int_0^\ell f^*g''\,dz
       +[f^*g']_0^\ell.
\end{equation}
The differential operator $-d^2/dz^2$ thus requires more regularity
and a choice of boundary conditions.  For example, the closed form
\eqref{eq:form-h1} on $H^1(0,\ell)$ is associated with the Neumann
Laplacian, whose operator domain is smaller than $H^1$.
The form itself does not require us to apply a second derivative to
every admissible function.  More abstractly, a continuous form on a
space $V$ defines a map $V\to V^*$, which need not map all of $V$
into $\mathcal H$.

\subsection{One-dimensional outgoing boundary form}
\label{sec:forms-1d}
The one-dimensional bath of Sec.~\ref{sec:oneD} offers an elementary example of the same boundary-form mechanism. Consider the interval $0\le z\le \ell$ at positive frequency, with $k=\omega/c$ and Maxwell wave expression
\begin{equation}
 L_{0,\mathrm{1D}}=\epsilon_0c^2(\partial_z^2+k^2).
\end{equation}
For smooth test functions $f,g$, integration by parts yields
\begin{align}
 \langle f,L_{0,\mathrm{1D}}g\rangle
 ={}&\epsilon_0c^2\int_0^\ell\!dz\,
 (k^2f^*g-f'^*g')\nonumber\\
 &+\epsilon_0c^2[f^*g']_0^\ell.
\end{align}
An outgoing wave satisfies $g'(\ell)=ikg(\ell)$ at the right endpoint and $g'(0)=-ikg(0)$ at the left endpoint. Equivalently, its outward normal derivative obeys $\partial_n g=ikg$ at both ends. Consequently, the outgoing boundary contribution is
\begin{equation}\label{eq:forms-1d-out}
 b_{\mathrm{out}}(f,g)=i\epsilon_0c^2k
 [f^*(0)g(0)+f^*(\ell)g(\ell)].
\end{equation}
The incoming condition reverses the sign of this term. The advanced-minus-retarded half difference is therefore
\begin{equation}\label{eq:forms-1d-pi}
 \frac12(b_{\mathrm{out}}^\dagger-b_{\mathrm{out}})(f,g)
 =-i\epsilon_0c^2k[f^*(0)g(0)+f^*(\ell)g(\ell)].
\end{equation}
The coefficient per endpoint is $-i\epsilon_0c\omega$, in agreement with Eq.~\eqref{eq:1d-bath}. In this boundary-form notation, the endpoint values are boundary values (called traces in Sobolev-space theory) of the test functions; they should not be confused with the point-sampled matrix normalization used in Sec.~\ref{sec:oneD}. This calculation is the scalar counterpart of the Maxwell surface form derived below: the boundary consists of two points rather than a sphere.

\subsection{The Maxwell form and Green's identity}
Let $\Omega$ be a bounded three-dimensional region containing the material objects, with boundary $\partial\Omega$. For electromagnetism, the corresponding first-derivative space is
\begin{equation}\label{eq:hcurl}
 H(\operatorname{curl},\Omega)=
 \{\mathbf f\in L^2(\Omega)^3:\nabla\times \mathbf f\in L^2(\Omega)^3\}.
\end{equation}
The natural boundary datum is the \emph{tangential trace} of the
field.  On sufficiently regular boundaries, it is meaningful in a
weak trace sense even when a pointwise boundary value is unavailable
\cite{Monk2003}.

At fixed real $\omega$, the differential expression underlying the Maxwell wave operator, as 
used in Sec.~II, is
\begin{equation}\label{eq:forms-maxwell}
 L_0=\epsilon_0\omega^2I
       -\mu_0^{-1}\nabla\times\nabla\times.
\end{equation}
Its interior volume form is
\begin{align}\label{eq:forms-volume}
 a_0(\mathbf f,\mathbf g)={}&\epsilon_0\omega^2
        \int_\Omega \mathbf f^*\cdot \mathbf g\,dV\nonumber\\
 &-\frac1{\mu_0}\int_\Omega
 (\nabla\times \mathbf f)^*\cdot(\nabla\times \mathbf g)\,dV.
\end{align}
This form is Hermitian for real $\omega$.  In particular, it does
not by itself distinguish waves entering from waves leaving the
region.

For smooth $\mathbf f,\mathbf g$, let $\hat{\mathbf n}$ denote the general outward boundary normal. Applying integration by parts to the curl--curl term gives the vector Green's identity
\begin{align}\label{eq:forms-green}
 \langle \mathbf f,L_0\mathbf g\rangle_\Omega
 =a_0(\mathbf f,\mathbf g)-\frac1{\mu_0}
 \oint_{\partial\Omega}dS\,
 \mathbf f^*\cdot[\hat{\mathbf n}\times(\nabla\times \mathbf g)].
\end{align}
The sign can be checked directly from
$\nabla\cdot(\mathbf f^*\times \mathbf F)
 =\mathbf{F}\cdot(\nabla\times \mathbf f^*)-\mathbf f^*\cdot(\nabla\times \mathbf{F})$.
Taking the difference of \eqref{eq:forms-green} and its adjoint
leaves only a boundary contribution.  Thus the bulk differential
expression is formally Hermitian, but the boundary-value problem
need not be self-adjoint.  The Hilbert-space adjoint of an unbounded
operator depends on its \emph{domain}, not only on the displayed
differential expression.
For clarity, write $L_{\rm out}$ and $L_{\rm in}$ for realizations
of the same interior Maxwell wave expression with outgoing and
incoming boundary prescriptions, respectively.  Their boundary
conditions are part of their domains.  We do not identify their
formal difference with a bulk differential operator: instead, we
compare the corresponding weak boundary forms on common admissible
tangential traces. These are weak formulations incorporating the
radiation condition, not unrestricted identities for arbitrary
$H(\operatorname{curl},\Omega)$ fields.  This is the continuum meaning intended below.

\subsection{Outgoing waves and the Silver--M\"uller condition}
Now enclose all matter in a sphere $S_R$, with outward normal
$\hat{\mathbf n}=\hat{\mathbf R}$, and take $R$ into the radiation zone.
For the $e^{-i\omega t}$ convention and $\omega>0$, the
Silver--M\"uller outgoing radiation condition~\cite{ColtonKress2013} is, to leading order,
\begin{equation}\label{eq:forms-sm}
 \mathbf H\simeq Z_0^{-1}\hat{\mathbf R}\times\mathbf E,
 \qquad Z_0=\mu_0c,
\end{equation}
where $Z_0$ is the vacuum impedance, $\mathbf E=i\omega\mathbf A$, and
$\mu_0\mathbf H=\nabla\times\mathbf A$.
Consequently,
\begin{align}\label{eq:forms-sm-curl}
 \hat{\mathbf R}\times(\nabla\times\mathbf A)
 &\simeq ik_0\hat{\mathbf R}\times(\hat{\mathbf R}\times\mathbf A)\nonumber\\
 &=-ik_0P_R\mathbf A,\qquad
 P_R=I-\hat{\mathbf R}\hat{\mathbf R}^T,
\end{align}
with $k_0=\omega/c$.  The projector $P_R$ removes the normal
component, leaving the two tangential radiation polarizations.
The approximation becomes asymptotically exact for radiation from
sources confined to a finite region; terms of relative order
$1/(k_0R)$ and the exact finite-sphere nonlocal response are not
included.

Insert \eqref{eq:forms-sm-curl} into the boundary term of
\eqref{eq:forms-green}.  The resulting \emph{outgoing inverse
boundary form} is
\begin{align}\label{eq:forms-out}
 a_{\mathrm{out}}(\mathbf f,\mathbf g)
 =a_0(\mathbf f,\mathbf g)+i\epsilon_0c\omega
 \oint_{S_R}dS\,\mathbf f^*\cdot P_R\mathbf g.
\end{align}
Its adjoint has the opposite sign of the imaginary boundary term:
\begin{align}\label{eq:forms-in}
 a_{\mathrm{out}}^\dagger(\mathbf f,\mathbf g)
 =a_0(\mathbf f,\mathbf g)-i\epsilon_0c\omega
 \oint_{S_R}dS\,\mathbf f^*\cdot P_R\mathbf g.
\end{align}
Hence
\begin{align}\label{eq:forms-pi}
 \frac12(a_{\mathrm{out}}^\dagger-a_{\mathrm{out}})(\mathbf f,\mathbf g)
 =-i\epsilon_0c\omega
 \oint_{S_R}dS\,\mathbf f^*\cdot P_R\mathbf g.
\end{align}
Comparing with the definition
$\Pi_\infty^r=\tfrac12[(v^a)^{-1}-(v^r)^{-1}]$,
we identify the radiative self-energy \emph{per unit area},
\begin{equation}\label{eq:forms-sphere-pi}
 \boxed{\Pi_{\infty,S_R}^r
 =-i\epsilon_0c\omega(I-\hat{\mathbf R}\hat{\mathbf R}^T).}
\end{equation}
The sign can be checked without relying on the abstract form
notation.  For a locally outgoing transverse plane wave, the
Silver--M\"uller relation gives
$\hat{\mathbf R}\times(\nabla\times\mathbf A)
=-ik_0P_R\mathbf A$.  The minus sign in the boundary term of
Eq.~\eqref{eq:forms-green} then produces
$+i\epsilon_0c\omega P_R$ in the outgoing weak inverse form.
The incoming relation reverses this sign, so their
advanced-minus-retarded half difference is
$-i\epsilon_0c\omega P_R$.  For $\omega>0$ the resulting
broadening is
\begin{equation}\label{eq:forms-positive-broadening}
\Gamma_{\infty,S_R}
=i(\Pi_{\infty,S_R}^r-\Pi_{\infty,S_R}^a)
=2\epsilon_0c\omega P_R\succeq0.
\end{equation}
This is the expected sign for a passive radiation channel.

\subsection{Meaning and limitations of the inverse formula}
Equation~\eqref{eq:forms-pi} is an asymptotic boundary \emph{form} identity. 
It should not be read as an unrestricted subtraction of two
second-order differential expressions on the same unspecified
domain.  For an open continuum problem, outgoing and incoming
realizations have different boundary prescriptions; their adjoints
are defined only after these prescriptions and the relevant spaces
are specified.  The boundary form is a practical way to compare
them on common admissible test fields; this comparison is not a claim
that the unbounded operators can be subtracted on arbitrary domains.  Our finite sampled-matrix
inverse, by contrast, has an ordinary matrix adjoint after a
specific contact-term regularization.  
The two methods are complementary; 
their agreement does not imply that every discretization is identical to the continuum construction.

This derivation also explains why the planar and spherical baths
both act on tangential boundary variables.  For a sphere in the far
field, the wave vector is asymptotically normal to the surface, so
$P_R$ follows directly from the radiation condition.  In the planar
sampled inverse, the same two independent boundary channels emerge
only after treating the longitudinal contact term.  The
Silver--M\"uller argument is a fourth route to the bath at infinity,
alongside exterior dust, lead elimination, and sampled Green's
function inversion.

\section{Spherical bath in the radiation zone}\label{sec:sphere}
Section~\ref{sec:forms} derived the spherical bath coefficient from the Silver--M\"uller radiation condition and a boundary sesquilinear form. Here we check it independently using the far-field Green's function and the exterior-dust construction.
The spherical boundary construction was used previously in photon NEGF transport calculations \cite{ZhangZhuZhangWang2022} and developed systematically in Ref.~\cite{TangRenWang2026}.
Consider finite material objects enclosed by a sphere $S_R$ of radius $R$, with $R$ much larger than both the object size and the relevant wavelengths. For observation points in a bounded central region, the outgoing vacuum vector-potential Green tensor has the asymptotic form
\begin{equation}\label{eq:sphere-free}
v^r(\mathbf r,R\hat{\mathbf R})\simeq
-\frac{e^{ik_0R-ik_0\hat{\mathbf R}\cdot\mathbf r}}
{4\pi\epsilon_0c^2R}\,P_R,\quad
P_R=I-\hat{\mathbf R}\hat{\mathbf R}^T,
\end{equation}
for $\omega>0$, where $k_0=\omega/c$. The projector is transverse to the outgoing direction, and $P_R^2=P_R$.

\explanatoryheading{Why a sphere is the natural boundary at infinity.}
For a finite collection of objects, every far-field observation direction can be labeled by a unit vector $\hat{\mathbf R}$.  At a sufficiently large radius, the electromagnetic field is locally a transverse plane wave, so the two allowed field components lie in the plane perpendicular to $\hat{\mathbf R}$.  The matrix $P_R=I-\hat{\mathbf R}\hat{\mathbf R}^T$ enforces precisely this geometrical condition.  It is a projector: it removes the radial component and leaves both transverse components unchanged.

The factor $1/R$ in the outgoing Green's function might initially suggest that a bath moved infinitely far away would have no effect.  However, the surface area grows as $R^2$.  In a field correlation, two propagators contribute $1/R^2$, and the surface integration supplies the compensating $R^2$.  Thus the boundary contribution has a finite large-radius limit.  This is the spherical counterpart of the compensation between weak local absorption and an infinite integration range in one dimension.

\subsection{Volume dust reduced to a surface}
Let $s$ denote distance outward from the spherical surface. With a weak exterior conductivity $\sigma=2\epsilon_0\eta$, outgoing radiation is attenuated by $e^{-\eta s/c}$ in field amplitude. The vacuum Green's function product scales as $R^{-2}$, while the radial volume element supplies $R^2ds\,d\Omega$. Combining Eq.~(\ref{eq:dust-pi}) with Eq.~(\ref{eq:dust-radial}) yields the effective surface spectral operator
\begin{equation}
\mathcal Q_{\infty,S_R}=-2i\epsilon_0c\omega P_R.
\end{equation}
Thus the local radiative bath self-energy, \emph{per unit surface area}, is
\begin{equation}\label{eq:sphere-bath}
\Pi_{\infty,S_R}^r=-i\epsilon_0c\omega
\left(I-\hat{\mathbf R}\hat{\mathbf R}^T\right).
\end{equation}
The surface expression is understood inside an integral $\int_{S_R}dS$ and in the limit of a radiation-zone sphere. It is not the exact Dirichlet-to-Neumann operator on an arbitrary finite sphere: the latter is generally nonlocal and includes reactive terms.

\explanatoryheading{What the angular average tests.}
The bath is locally transverse to each surface normal, but the observation point receives waves from every direction.  Rotational symmetry then forces the angularly averaged correlation to be proportional to the three-dimensional identity matrix.  The coefficient follows from $\operatorname{Tr}P_R=2$: each propagation direction has exactly two radiative polarizations.  The angular integral below therefore checks simultaneously the tensor structure, the polarization count, and the surface normalization.

\subsection{Consistency with the free spectral function}
At the center, using $dS=R^2d\Omega$ and Eq.~(\ref{eq:sphere-free}), the two-bath-factor identity gives
\begin{align}
&\int_{S_R}dS\,v^r(0,R\hat{\mathbf R})\,
(\Pi_\infty^r-\Pi_\infty^a)
\,v^a(R\hat{\mathbf R},0)\nonumber\\
&\qquad=-\frac{i\omega}{8\pi^2\epsilon_0c^3}
\int d\Omega\,P_R
=-\frac{i\omega}{3\pi\epsilon_0c^3}I,
\label{eq:sphere-spectral}
\end{align}
where $\int d\Omega\,P_R=(8\pi/3)I$, following from rotational symmetry and $\int d\Omega\,\hat R_i\hat R_j=(4\pi/3)\delta_{ij}$. This agrees with $v^r(0,0)-v^a(0,0)$ in free space for the radiative transverse sector. The equality provides a normalization test for the factor $\epsilon_0c\omega$ and the two transverse polarizations.

\section{Physical consistency checks}\label{sec:checks}

\explanatoryheading{Why a blackbody test is more than dimensional analysis.}
A radiation bath must satisfy two requirements: it must let photons escape, and it must supply the fluctuations of thermal radiation when its temperature is nonzero.  An outgoing boundary condition tests only the first requirement.  The Planck-spectrum calculation tests the second, including the absolute coefficient and the two transverse polarizations.  In empty space, with all external baths at the same temperature, the Keldysh construction must reproduce the free thermal field correlation without reference to the arbitrary location of the enclosing boundary.

The checks below play different roles.  Planck's law and the
Stefan--Boltzmann limit test thermal fluctuations and absolute
normalization, while the pressure comparison tests the tensor
structure of the electric and magnetic correlations.  The
conservation identities in Sec.~\ref{sec:conservation} instead
verify algebraic compatibility of the Green's functions and
reservoir self-energies under their stated assumptions.

\subsection{Planck's law from the spherical bath}\label{sec:planck-normal-order}
For a thermal radiation environment, $D^<=N_\infty(D^r-D^a)$ in empty space. From Eq.~(\ref{eq:sphere-spectral}),
\begin{equation}
i\,\mathrm{Tr}\,D^<(0,0;\omega)
=\frac{\omega}{\pi\epsilon_0c^3}N_\infty(\omega),\qquad\omega>0.
\end{equation}
The positive-frequency lesser function alone accounts for one operator ordering. For a real field, the positive- and negative-frequency contributions enter the equal-time variance with the two corresponding operator orderings. After subtracting the zero-point part, the normally ordered thermal variance has twice the positive-frequency lesser contribution. Thus
\begin{equation}\label{eq:planck-electric}
\epsilon_0\langle:\mathbf E^2:\rangle
=\int_0^\infty d\omega\,\frac{\hbar\omega^3}{\pi^2c^3}N_\infty(\omega).
\end{equation}
The magnetic contribution is equal in free radiation, $\langle:\mathbf B^2:\rangle/\mu_0=\epsilon_0\langle:\mathbf E^2:\rangle$. Thus
\begin{equation}\label{eq:planck-total}
\boxed{u_T=\int_0^\infty d\omega\,\frac{\omega^2}{\pi^2c^3}\,\hbar\omega N_\infty(\omega),}
\end{equation}
which is Planck's thermal energy density. The vacuum zero-point part is excluded by normal ordering; it must be retained in observables for which vacuum stress is physically relevant.
\subsection{Planar blackbody radiation}
A second check places equal-temperature planar baths at $z=\pm L$, with no material between them. The Keldysh equation gives
\begin{equation}
D^<(0,0)=\sum_{s=\pm}v^r(0,sL)\Pi_s^<v^a(sL,0).
\end{equation}
For propagating channels the propagation phases cancel in the products. The two tangential boundary baths supply both propagation directions. When the full three-component propagation tensor is retained, the normal electric-field component of TM waves is reconstructed even though the bath matrix itself is tangential. This reproduces the free-space equilibrium FDT and therefore Eq.~(\ref{eq:planck-electric}). The equality assumes the boundary planes lie in homogeneous vacuum and the positive-frequency outgoing branch is used consistently.
\paragraph{Explicit angular integration.}\label{sec:planar-blackbody-detail}
The planar test can be evaluated without invoking the spherical result.
For a fixed positive frequency, let $k_0=\omega/c$ and write the propagating
transverse wave vector as $q_\perp=k_0\sin\theta$,
$k=k_0\cos\theta$, $0\le\theta<\pi/2$.  The \emph{tangential} trace of the free dyadic at coincident
positions is
\begin{equation}\label{eq:planar-spectral-trace}
 i\,\mathrm{Tr}_{\parallel}\,[v^r-v^a]
 =\frac{1}{\epsilon_0c^2 k}
 \left(1+\frac{k^2}{k_0^2}\right),
\end{equation}
for $q_\perp<k_0$. The normal $zz$ component adds
$q_\perp^2/(\epsilon_0c^2 k k_0^2)$, so that the full
three-dimensional trace is $2/(\epsilon_0c^2k)$, as
required for two physical photon polarizations.
With the Fourier measure fixed in Sec.~\ref{sec:conventions},
\begin{align}
 &\int_{q_\perp<k_0}\frac{d^2q_\perp}{(2\pi)^2}
 \frac{1+k^2/k_0^2}{k}
 =\frac{k_0}{2\pi}\int_0^1
 \frac{x(2-x^2)}{\sqrt{1-x^2}}\,dx
 =\frac{2k_0}{3\pi},\nonumber\\
 &\int_{q_\perp<k_0}\frac{d^2q_\perp}{(2\pi)^2}
 \frac{q_\perp^2/k_0^2}{k}=\frac{k_0}{3\pi}.
\end{align}
The first dimensionless integral is $4/3$; the sum of the tangential and normal contributions is $k_0/\pi$. Multiplication by the electric-field factor
$\epsilon_0\hbar\omega^2$ and the appropriate equal-time thermal
occupation recovers Eq.~(\ref{eq:planck-electric}).  This exercise is
useful for checking both polarization channels and the
$z$ component of the electric field; retaining only the tangential
trace would give an incorrect intermediate angular factor.

\subsection{Stefan--Boltzmann limit}
The isotropic blackbody intensity gives an outward energy flux $J=(c/4)u_T$. Integrating Eq.~(\ref{eq:planck-total}) and using
\begin{equation}
\int_0^\infty dx\,\frac{x^3}{e^x-1}=\frac{\pi^4}{15},
\end{equation}
we obtain
\begin{equation}\label{eq:stefan}
J=\sigma_{\rm SB}T^4,\qquad
\sigma_{\rm SB}=\frac{\pi^2k_B^4}{60\hbar^3c^2}.
\end{equation}
Two ideal blackbody reservoirs at temperatures $T_L,T_R$ exchange net power per area $\sigma_{\rm SB}(T_L^4-T_R^4)$, with the sign chosen from left to right. This provides an additional integrated normalization check, following from the Planck spectrum and the isotropic-flux relation rather than an independent derivation of the bath.
\subsection{Radiated power: Meir--Wingreen current and Poynting flux}
As a further check, let the bath at infinity be at zero temperature,
$N_\infty(\omega)=0$ for $\omega>0$, and consider stationary radiation
emitted by matter enclosed by a large sphere $S_R$.  For positive frequencies, the bath has
$\Pi_\infty^<=0$ and $\Pi_\infty^>=-i\Gamma_\infty$, with the
radiation-zone surface broadening
$\Gamma_{\infty,S_R}=2\epsilon_0c\omega P_R$,
$P_R=I-\hat{\mathbf R}\hat{\mathbf R}^{T}$.
The Meir--Wingreen current, positive when the bath emits into the field,
therefore gives the outward power
\begin{align}
 -I_\infty
 &=\int_0^\infty\!\frac{d\omega}{2\pi}\,\hbar\omega\,
   \mathrm{Tr}[i\Gamma_\infty D^<]\label{eq:poynting-mw}\\
 &=2\epsilon_0c\hbar\int_0^\infty\!\frac{d\omega}{2\pi}\,
   \omega^2\oint_{S_R}\!dS\,
   i\,\mathrm{Tr}[P_RD^<(\mathbf R,\mathbf R;\omega)].
\nonumber
\end{align}
On the other hand, the outgoing Silver--M\"uller condition gives
$\mathbf H \simeq \epsilon_0c\,\hat{\mathbf R}\times\mathbf E$ in the radiation
zone. Hence the normally ordered
Poynting flux is
\begin{align}
 P_{\rm Poynting}
 &=\oint_{S_R}\!dS\,\hat{\mathbf R}\cdot
   \langle:\mathbf E\times\mathbf H:\rangle\nonumber\\
 &=\epsilon_0c\oint_{S_R}\!dS\,
   \langle:\mathbf E\cdot P_R\mathbf E:\rangle
 =-I_\infty\label{eq:poynting-flux}.
\end{align}
In the last equality we used $\mathbf E=-\partial_t\mathbf A$ and
the positive-frequency normal-ordering prescription explained
in Sec.~\ref{sec:planck-normal-order}, which gives twice the positive-frequency
lesser contribution to the electric-field variance.  The bath current and outward Poynting flux therefore agree exactly, including their sign and numerical prefactors.  The equality applies to the outgoing radiation from
the matter when the incoming bath occupation vanishes; vacuum terms
carry no net normally ordered flux.

\subsection{Radiation pressure: scattering-theory benchmark}
\label{sec:sheet-pressure}

We consider a planar material sheet at $z=0$, with vacuum extending to
$z\to\pm\infty$.  The left and right exterior half-spaces are
represented by photonic baths at infinity.  A monochromatic plane wave
of frequency $\omega$, transverse wave vector $\mathbf q_\perp$,
and TE or TM polarization is incident from the left.  The sheet may
reflect, transmit, or absorb the incident radiation.  We determine its
force per unit area from the Maxwell stresses on the two sides.

The electromagnetic stress tensor is \cite{Jackson1999}
\begin{equation}
T_{ij}=\epsilon_0\left(E_iE_j-\tfrac12\delta_{ij}E^2\right)
+\mu_0^{-1}\left(B_iB_j-\tfrac12\delta_{ij}B^2\right).
\end{equation}
For an isotropic blackbody field, $\langle T_{ij}\rangle=-u_T\delta_{ij}/3$ in this stress convention. For a single normally incident plane wave, the momentum flux is its energy flux divided by $c$. These provide limiting tests of a Green's function stress calculation. The explicit planar TE/TM stress calculation and its agreement with scattering theory are given below. The extension to several independent incident and emitting sources is organized in Appendix~\ref{app:stress-sources}.

For a plane wave incident from the left at angle $\theta$ in vacuum,
let $J_{\rm in}$ be the \emph{normal} incident energy flux per unit
surface area, and let $R_\sigma$, $T_\sigma$, and $A_\sigma$
be the energy reflection, transmission, and absorption probabilities
for polarization $\sigma=s,p$.  Assume a stationary passive sheet,
identical vacuum on both sides, and no frequency conversion.
Then $R_\sigma+T_\sigma+A_\sigma=1$.
The incoming photon has normal momentum $\hbar k=\hbar\omega\cos\theta/c$.
The reflected photon reverses its normal momentum, while the
transmitted photon retains it.  The force per unit area exerted on the
sheet by that incident channel is therefore
\begin{align}
 \frac{F_{z,\sigma}}{\mathcal A}
 &=\frac{J_{\rm in,\sigma}}{c}\cos\theta
 \left(1+R_\sigma-T_\sigma\right)\nonumber\\
 &=\frac{J_{\rm in,\sigma}}{c}\cos\theta
 \left(A_\sigma+2R_\sigma\right)\label{eq:sheet-pressure-single}.
\end{align}
At normal incidence a perfectly absorbing sheet receives $J_{\rm in}/c$,
while a perfectly reflecting one receives $2J_{\rm in}/c$.
For oblique incidence the factor $\cos\theta$ is present because
$J_{\rm in}$ is defined per unit area of the sheet, rather than
per unit area perpendicular to the incident ray.

For a general radiation environment, the force must be found by
subtracting the momentum fluxes on the two sides of the sheet,
including radiation incident from both directions and any emitted
radiation.  Equation~(\ref{eq:sheet-pressure-single}) is a
scattering-theory benchmark for that calculation, not a replacement
for a complete Maxwell-stress evaluation in an absorbing,
nonequilibrium system.  In particular, vacuum-stress terms relevant
to Casimir forces are not removed by the normal ordering appropriate
to the thermal Planck-spectrum calculation.

\subsection{Maxwell stress from the photon lesser Green's function}
\label{sec:lesser-pressure}
We now verify the radiation-pressure result of
Sec.~\ref{sec:sheet-pressure} using the bath-at-infinity
self-energy. We consider radiation incident from the left
on a stationary, linear sheet, with no incoming population
from the right bath. The sheet's own fluctuations are
excluded from this single-sided comparison.

For a propagating channel of polarization
$\sigma=s,p$, the left bath has
\begin{equation}
\Pi_{L,\sigma}^r=-i\omega Y_\sigma,\quad
\Gamma_{L,\sigma}=2\omega Y_\sigma,\quad
\Pi_{L,\sigma}^<=-iN_L\Gamma_{L,\sigma},
\end{equation}
where $Y_\sigma$ is the electromagnetic wave admittance, and $N_L$ is the Bose function at the temperature of 
the left bath. 

For radiation incident only from the left, the bath-generated contribution to the lesser Green's function is
\begin{equation}
D_{L,\sigma}^<=D_\sigma^r\Pi_{L,\sigma}^<D_\sigma^a.
\end{equation}
In the coordinate representation, using $\Pi_{L,\sigma}^<=-iN_L\Gamma_{L,\sigma}$, this becomes
\begin{equation}
i\hbar D_{L,\sigma}^<(z,z')
=\hbar N_L\Gamma_{L,\sigma}
D_\sigma^r(z,z_L)D_\sigma^a(z_L,z'),
\end{equation} 
where $z_L$ denotes the left bath boundary. The dressed propagator can be expressed in terms of the scattering profile as
\begin{equation}
D_\sigma^r(z,z_L)=g_\sigma^r e^{-ikz_L} f_\sigma(z),
\qquad
g_\sigma^r=-\frac{i}{2\omega Y_\sigma}.
\end{equation}
Here \(z_L<0\), and the expression applies to observation points \(z>z_L\). The propagation phase cancels in the correlation function.
Since $D_\sigma^a(z_L,z')=[D_\sigma^r(z',z_L)]^*$ and $\Gamma_{L,\sigma}=2\omega Y_\sigma$, we obtain
\begin{equation}
i\hbar D_{L,\sigma}^<(z,z')
=\frac{\hbar N_L}{2\omega Y_\sigma}
f_\sigma(z)f_\sigma^*(z'),
\end{equation}
where
\[
f_\sigma(z)=
\begin{cases}
e^{ikz}+\eta_\sigma r_\sigma e^{-ikz},&z<0,\\
t_\sigma e^{ikz},&z>0,
\end{cases}
\quad\eta_s=1,\:\eta_p=-1.
\]
Here $r_\sigma$
and $t_\sigma$ refer to total electric-field amplitudes
in the polarization basis used below. Thus, the incoming,
reflected, and transmitted field correlations are
generated by the bath lesser self-energy and the
dressed retarded Green's function.

The electric and magnetic correlations follow from
$\mathbf E=-\partial_t\mathbf A$ and
$\mathbf B=\nabla\times\mathbf A$. For a channel with
$k_0=\omega/c$, $q=|\mathbf q_\perp|$, and
$k=(k_0^2-q^2)^{1/2}>0$, choose
$\mathbf q_\perp=q\hat{\mathbf x}$.
For TE polarization, the electric field is along
$\hat{\mathbf y}$; for TM polarization, use
\begin{equation}
\mathbf e_{p\pm}
=\frac{\pm k\hat{\mathbf x}-q\hat{\mathbf z}}{k_0}.
\end{equation}
Evaluating the Maxwell stress tensor gives, for either
polarization,
\begin{equation}
\overline T_{zz}^{(\sigma)}
=-\frac{\epsilon_0}{2}\frac{k^2}{k_0^2}
\left(\langle|a|^2\rangle+\langle|b|^2\rangle\right),
\end{equation}
where $a$ and $b$ are the forward and backward
total electric-field amplitudes obtained from the
bath-generated covariance. The interference terms
cancel, and the normal electric-field component
must be included for TM polarization.

For radiation incident only from the left,
$(a_L,b_L)=(A,r_\sigma A)$ and
$(a_R,b_R)=(t_\sigma A,0)$.
The stress difference across the sheet yields
\begin{equation}
\frac{F_{z,\sigma}}{\mathcal A}
=\frac{\epsilon_0}{2}\frac{k^2}{k_0^2}
\langle|A|^2\rangle
(1+R_\sigma-T_\sigma).
\end{equation}
Using the incident normal energy flux
$J_{\mathrm{in},\sigma}
=\epsilon_0 c(k/k_0)\langle|A|^2\rangle/2$,
we obtain
\begin{equation}
\boxed{
\frac{F_{z,\sigma}}{\mathcal A}
=\frac{J_{\mathrm{in},\sigma}}{c}
\frac{k}{k_0}(1+R_\sigma-T_\sigma).
}
\end{equation}
This agrees with Eq.~(\ref{eq:sheet-pressure-single})
and verifies that the bath-generated field
correlations reproduce the radiation pressure.
To calculate the complete nonequilibrium force, fluctuations from both exterior baths and the material sheet must be included.

\section{Spectral identities and energy conservation}\label{sec:conservation}
The relation between the radiation bath and the optical theorem is most transparent when we distinguish the free inverse operator from material polarization. From the Dyson equation,
\begin{equation}
(D^r)^{-1}=(v^r)^{-1}-\Pi_{\rm mat}^r,
\end{equation}
and its adjoint, the resolvent identity gives
\begin{align}\label{eq:optical-full}
D^r-D^a
&=D^r\big[(v^a)^{-1}-(v^r)^{-1}
+\Pi_{\rm mat}^r-\Pi_{\rm mat}^a\big]D^a\nonumber\\
&=D^r\big[\Pi_\infty^r-\Pi_\infty^a
+\Pi_{\rm mat}^r-\Pi_{\rm mat}^a\big]D^a.
\end{align}
In the second line the notation $\Pi_\infty^r-\Pi_\infty^a=(v^a)^{-1}-(v^r)^{-1}$ is used. This is an identity for the chosen operator representation; in a continuum setting it includes the radiation boundary prescription. Omitting the environmental spectral term generally destroys the equality even when the material absorption is zero.

\explanatoryheading{The missing term in a material-only optical theorem.}
The retarded free Green's function has a nonzero spectral density even in empty space.  If one were to omit the radiation contribution on the right-hand side of Eq.~(\ref{eq:optical-full}), setting the material polarization to zero would predict $v^r-v^a=0$, which is false for propagating vacuum modes.  This limiting case gives a quick diagnostic of the correct optical theorem.  The bath-at-infinity term restores the free-space spectral density while leaving the material Dyson equation unchanged.

\subsection{Thermal equilibrium}
If all material objects and the incoming radiation environment have the same Bose function $N$, their lesser polarizations satisfy $\Pi_\alpha^<=N(\Pi_\alpha^r-\Pi_\alpha^a)$ and similarly for $\Pi_\infty^<$. Combining Eq.~(\ref{eq:Keldysh-reservoir}) with Eq.~(\ref{eq:optical-full}) gives
\begin{equation}
D^<=N(D^r-D^a),
\end{equation}
which is the global fluctuation--dissipation theorem. This is the most direct equilibrium consistency check of the bath construction.
\subsection{Energy currents and reservoir bookkeeping}
The photon Meir--Wingreen expression for a material object $\alpha$ (by analogy with the electronic transport formula of Ref.~\cite{MeirWingreen1992}) can be written, with a specified sign convention for positive energy flow, in terms of $D^{</>}$ and $\Pi_\alpha^{</>}$. When every material subsystem is in local thermal equilibrium, it reduces to a Landauer-type sum of spectral transmissions between reservoirs \cite{Landauer1957}. In that sum the radiation bath must be included alongside the material objects. Otherwise radiated energy leaving the modeled region is absent from the reservoir balance.
The resulting conservation statement is
\begin{equation}\label{eq:energy-balance}
I_\infty+\sum_{\alpha\in\mathrm{matter}}I_\alpha=0
\end{equation}
for a stationary system without an external power source, provided the material self-energies and current definitions satisfy the corresponding conserving identities. The optical theorem is a necessary spectral ingredient, but it does not by itself prove that an arbitrary approximation for $\Pi_{\rm mat}$ conserves energy.
The force and torque formulas contain spatial and rotational generators acting on Green's functions. They do not generally admit the same simple Landauer difference-of-occupations form as the energy current. We therefore keep their conservation analysis distinct from Eq.~(\ref{eq:energy-balance}).

\explanatoryheading{What is meant by conservation here?}
The equality $\sum_\lambda I_\lambda=0$ expresses stationary bookkeeping: energy leaving one reservoir enters the photon field and ultimately another reservoir, including the exterior radiation bath.  It does not mean that every object carries zero current.  Nor does it assert that an arbitrary approximate polarization automatically conserves energy.  The following cancellation relies on compatible lesser, greater, retarded, and advanced components, together with a well-defined trace.  Stating these premises is especially important for students who may later use approximate electron--photon self-energies.

\subsection{Algebraic current cancellation in a stationary system}
\label{sec:current-cancellation}
For clarity, consider the trace over a finite set of photon degrees of
freedom, or a continuum representation in which the traces and
boundary operations are well defined.  Include the environmental bath
as one more reservoir, and put
$\Pi^{</>}_{\rm tot}=\sum_{\lambda}\Pi^{</>}_\lambda$,
where $\lambda$ runs over all material reservoirs and the bath at infinity.
The reservoir-resolved spectral energy current may be defined by
\begin{equation}\label{eq:current-lesser-greater}
 I_\lambda=\int_0^\infty\frac{d\omega}{2\pi}\,
 \hbar\omega\,\mathrm{Tr}
 \big[\Pi_\lambda^>D^<-\Pi_\lambda^<D^>\big],
\end{equation}
with positive current directed \emph{into the photon field}.
This formula assumes the chosen normal-ordering and positive-frequency
convention; the equivalent full-frequency form must be used if
frequency mixing is present.
For a stationary Gaussian field with the exact Keldysh relations,
\begin{equation}
 D^{</>}=D^r\Pi^{</>}_{\rm tot}D^a,
 \qquad D^>-D^<=D^r-D^a,
\end{equation}
and the compatible optical theorem,
$D^r-D^a=D^r(\Pi^r_{\rm tot}-\Pi^a_{\rm tot})D^a$,
we find, frequency by frequency,
\begin{align}
& \sum_\lambda\mathrm{Tr}
 [\Pi_\lambda^>D^<-\Pi_\lambda^<D^>] \nonumber \\
& =\mathrm{Tr}[(\Pi_{\rm tot}^>-\Pi_{\rm tot}^<)D^<
 -\Pi_{\rm tot}^<(D^>-D^<)]\nonumber\\
& =\mathrm{Tr}[(\Pi_{\rm tot}^r-\Pi_{\rm tot}^a)D^<
 -\Pi_{\rm tot}^<(D^r-D^a)].
\end{align}
The last line vanishes by the two equivalent forms of the resolvent
identity, $D^r-D^a=D^r\Delta\Pi D^a=D^a\Delta\Pi D^r$,
where $\Delta\Pi=\Pi^r_{\rm tot}-\Pi^a_{\rm tot}$, and cyclicity
of the trace.  Hence
\begin{equation}
 \sum_\lambda I_\lambda=0.
\end{equation}
This derivation assumes that the greater/lesser components and
retarded self-energies are mutually consistent.  A nonconserving
approximation for a nonlinear or interacting material polarization
need not obey the premises.

\subsection{Landauer reduction and transmission symmetry}
For locally equilibrated reservoirs,
$\Pi_\lambda^<=-iN_\lambda\Gamma_\lambda$ and
$\Pi_\lambda^>=-i(1+N_\lambda)\Gamma_\lambda$,
where $\Gamma_\lambda=i(\Pi_\lambda^r-\Pi_\lambda^a)$.
Insertion into Eq.~(\ref{eq:current-lesser-greater}) gives
\begin{equation}\label{eq:landauer-bath}
 I_\lambda=\int_0^\infty\frac{d\omega}{2\pi}\,\hbar\omega
 \sum_\rho\mathcal T_{\rho\lambda}(\omega)
 [N_\lambda-N_\rho],
\end{equation}
with
\begin{equation}
 \mathcal T_{\rho\lambda}
 =\mathrm{Tr}[ D^r\Gamma_\rho D^a\Gamma_\lambda].
\end{equation}
The sign convention is the same as in the photon-transport chapter of Ref.~\cite{WangXuZhouInPreparation}: positive $I_\lambda$ denotes net power emitted by reservoir $\lambda$. 
Thus the occupation difference $N_\lambda-N_\rho$ is weighted by $\mathcal T_{\rho\lambda}$; for nonreciprocal systems the reversed coefficient need not be equal. The optical theorem supplies the row--column sum rule
\begin{equation}
 \sum_\rho\mathcal T_{\lambda\rho}=\sum_\rho\mathcal T_{\rho\lambda},
\end{equation}
which guarantees $\sum_\lambda I_\lambda = 0$ even when individual pairwise transmissions differ. 
This follows by tracing $\Gamma_\lambda$ times the two equivalent resolvent identities and using cyclicity.
Thus the total current vanishes without assuming pairwise reciprocity.
In particular, omission of
$\lambda=\infty$ from the sum is generally incorrect for a system
that emits propagating radiation.

\section{Discussion: Connections Between the Four Approaches}\label{sec:discussion}
The four approaches developed here represent the same physical process: eliminating the infinite electromagnetic environment while retaining its influence on a finite region containing the material objects. To see their connections, consider a partition of the free Maxwell wave operator into an interior region $C$ and an exterior region $E$. In a suitable discretized representation, eliminating the exterior gives the embedding self-energy
\begin{equation}
\Sigma_E^r=V_{CE}g_E^rV_{EC},
\end{equation}
where $g_E^r$ is the retarded Green's function of the isolated exterior, and $V_{CE}$ and $V_{EC}$ describe its coupling to the interior. The anti-Hermitian part of $\Sigma_E^r$ represents the radiative contribution.

\subsection{Dust absorption and surface Green's functions}
To connect the environmental-dust method with the surface Green's function, introduce infinitesimal absorption in the exterior through $L_E^r=H_E+iW_\eta$, where $H_E$ is Hermitian and $W_\eta$ is positive semidefinite. The resolvent identity gives $g_E^r-g_E^a=-2ig_E^rW_\eta g_E^a$. Consequently,
\begin{align}
\Gamma_E&=i(\Sigma_E^r-\Sigma_E^a) \nonumber\\
&=2V_{CE}g_E^rW_\eta g_E^aV_{EC}.
\end{align}
The right-hand side describes absorption distributed throughout the exterior, while the left-hand side describes the same loss through the surface self-energy. Although $W_\eta\to0$, the integrated absorption can remain finite because the exterior is unbounded. Thus, the surface Green's function incorporates the radiative contribution obtained from the dust model in the limiting-absorption procedure.

\subsection{Sampled inverses and boundary forms}
The connection to the sampled-matrix method follows from the Schur complement,
\begin{equation}
(v_{CC}^r)^{-1}=L_C-\Sigma_E^r.
\end{equation}
Thus, the inverse of the free Green's function restricted to the interior contains the exterior self-energy. When $L_C$ is Hermitian, its anti-Hermitian part is determined entirely by the exterior response. This explains why the sampled inverse reveals the bath self-energy at the endpoints in our three-site calculation.

The sesquilinear-form approach provides a continuum counterpart to this elimination procedure. Instead of eliminating exterior variables in a matrix, Green's identity expresses their effect through a boundary form. The outgoing Silver--M\"uller radiation condition then gives the spherical bath self-energy in the radiation-zone limit. The finite-dimensional Schur complement and the continuum boundary form therefore describe the same exterior response in different representations.

These connections hold within the assumptions of the respective constructions. In particular, the sampled Cartesian inverse requires the specified longitudinal contact-term regularization, while an exact finite-boundary embedding operator generally contains reactive as well as radiative contributions. The bulk absorption of the dust model and the outgoing boundary response are alternative descriptions of the same radiation loss and must not be counted twice.

\section{Conclusions}\label{sec:conclusions}
We have presented four complementary approaches to deriving the photonic bath-at-infinity self-energy within nonequilibrium Green's function theory: environmental dust, surface Green's functions, sampled Green's-function inversion, and sesquilinear forms. Particular attention has been given to the full Cartesian matrix-inversion method, where the longitudinal contact term is essential for obtaining the planar bath self-energy.

The sesquilinear-form approach provides a mathematical interpretation of the radiation bath. Although the differential expression underlying the Maxwell wave operator is formally Hermitian, its outgoing and incoming realizations differ through their boundary conditions. Green's identity and the Silver--M\"uller radiation condition make this difference explicit as an anti-Hermitian boundary contribution, yielding the spherical bath self-energy. This clarifies the meaning of the Hermitian conjugate of the inverse free Green's function when outgoing radiation boundary conditions are imposed. These results establish the mathematical origin of the bath-at-infinity self-energy and clarify its role in connecting the NEGF reservoir description with electromagnetic radiation boundary conditions.

The resulting bath self-energies reproduce established electromagnetic results, including the Planck spectrum, the Stefan--Boltzmann law, and radiation pressure. The agreement between the Meir--Wingreen formulas and calculations based on the Poynting vector and Maxwell stress tensor provides a further consistency check. The bath at infinity also supplies the radiative contribution required for energy conservation in photon NEGF.

\begin{acknowledgments}
The author thanks Gaomin Tang and Yuhua Ren for careful reading of the manuscript. 
This paper is based on the author's handwritten research notes, with assistance from ChatGPT in organizing the material and preparing the manuscript. The author takes responsibility for the scientific content and its verification.
\end{acknowledgments}

\appendix
\newpage

\section{An $N$-plane inverse and its endpoint spectral part}\label{app:Ninverse}
Here we give the intermediate steps behind the endpoint localization result.
Set $u_j=\exp(ikd_j)$ and $M_{ij}=\exp(ik|z_i-z_j|)$ for
$z_1<\cdots<z_N$.  Introduce the lower-bidiagonal matrix $B$ with
$B_{11}=1$, $B_{jj}=1$ and $B_{j,j-1}=-u_{j-1}$ for $j\geq2$.
Direct multiplication gives
\begin{equation}
 BMB^T=\operatorname{diag}(1,1-u_1^2,\ldots,1-u_{N-1}^2).
\end{equation}
For example, the off-diagonal $(j,j-1)$ entry vanishes because
$M_{j,j-1}-u_{j-1}M_{j-1,j-1}=u_{j-1}-u_{j-1}=0$.
Consequently,
\begin{equation}
 M^{-1}=B^T\operatorname{diag}\left(1,\frac1{1-u_1^2},\ldots,
 \frac1{1-u_{N-1}^2}\right)B.
\end{equation}
This proves tridiagonality without guessing an inverse.  In particular,
\begin{align}
 (M^{-1})_{11}&=\frac1{1-u_1^2},\\
 (M^{-1})_{NN}&=\frac1{1-u_{N-1}^2},\\
 (M^{-1})_{j,j+1}&=-\frac{u_j}{1-u_j^2},\\
 (M^{-1})_{jj}&=\frac1{1-u_{j-1}^2}
 +\frac1{1-u_j^2}-1\quad (1<j<N).
\end{align}
For propagating channels with $k$ real, the identities
\begin{align}
 \frac1{1-e^{2ix}}&=\frac12+\frac{i}{2}\cot x,\\
 -\frac{e^{ix}}{1-e^{2ix}}&=\frac1{2i\sin x}
\end{align}
show that $g^{-1}M^{-1}$, with $g=-i/(2\epsilon_0c^2k)$,
has purely real off-diagonal entries.  Every interior diagonal
has imaginary part zero because its two fractions contribute
$1/2+1/2-1=0$ to the real part of $M^{-1}$.
At either endpoint, the real part of $M^{-1}$ is $1/2$, and hence
\begin{equation}
 \frac{(g^{-1}M^{-1})^\dagger-g^{-1}M^{-1}}2
 =-i\epsilon_0c^2k\left(|1\rangle\langle1|+|N\rangle\langle N|\right).
\end{equation}
The conclusion holds for nonsingular sampled matrices; isolated values
$\sin(kd_j)=0$ must be approached by continuity or with a small
retarded regulator.  The result does not require equal plane spacings.

\subsection{Why the full Cartesian contact term has a simple limit}
Let the full sampled dyadic be reordered into tangential and normal
variables, $v_\delta=\bigl(\begin{smallmatrix}A&B\\C&E_\delta\end{smallmatrix}\bigr)$,
with $E_\delta=\delta^{-1}C_0+E_0$ and $C_0=(\epsilon_0\omega^2)^{-1}I_N$.
For fixed nonzero $\omega$, distinct sampling positions, and nonsingular
$A$, the Neumann expansion yields
\begin{equation}
 E_\delta^{-1}=\delta C_0^{-1}-\delta^2 C_0^{-1}E_0C_0^{-1}+O(\delta^3).
\end{equation}
The Schur complement is therefore
$S_\delta=A-\delta BC_0^{-1}C+O(\delta^2)$ and
$S_\delta^{-1}=A^{-1}+O(\delta)$.  The off-diagonal blocks of
$v_\delta^{-1}$ are $O(\delta)$, as is its normal--normal block.
This proves the limiting matrix inverse used in the text, rather than
merely deleting the normal variables before inversion.
The limit is not uniform when the tangential matrix approaches a
singular point (notably $k=0$).  Nor does this point-sampling argument
establish the same result for arbitrary regularizations of a continuum
operator; those are separate mathematical questions.



\section{Surface Green's function and boundary bookkeeping}\label{app:surface}\label{app:chain-detail}
This appendix spells out the cancellation that can be obscured by
calling the raw embedding contribution the ``bath self-energy.''
The site-weighted discrete free inverse has diagonal
$\epsilon_0h\omega^2-2K$ and nearest-neighbor entries $K$, with
$K=\epsilon_0c^2/h$.  The retarded bulk dispersion is
\begin{equation}
 \epsilon_0h\omega^2-2K+K(\lambda+\lambda^{-1})=0,
 \qquad \lambda=e^{i\vartheta}.
\end{equation}
The semi-infinite surface Green's function $g_s$ satisfies
$g_s^{-1}=\epsilon_0h\omega^2-2K-K^2g_s$.
Setting $g_s=-\lambda/K$ gives
\begin{align}
 g_s^{-1}&=-K\lambda^{-1},\\
 \epsilon_0h\omega^2-2K-K^2g_s
 &=-K(\lambda+\lambda^{-1})+K\lambda
 =-K\lambda^{-1}.
\end{align}
Thus the outgoing root indeed solves the quadratic equation, and
$\Sigma_B=K^2g_s=-K\lambda$.
If the retained central endpoint has only one internal bond, its
unembedded diagonal is $\epsilon_0h\omega^2-K$.  The effective
endpoint inverse after eliminating the exterior is therefore
\begin{equation}
 L_{\rm end}^{\rm eff}=\epsilon_0h\omega^2-K-\Sigma_B
 =\epsilon_0h\omega^2+K(\lambda-1).
\end{equation}
At fixed $\omega>0$, $\vartheta=\omega h/c+O(h^3)$, so
\begin{align}
 K(\lambda-1)
 &=\frac{\epsilon_0c^2}{h}\left(i\frac{\omega h}{c}
 -\frac{\omega^2h^2}{2c^2}+O(h^3)\right)\nonumber\\
 &=i\epsilon_0c\omega-\frac12\epsilon_0\omega^2h+O(h^2).
\end{align}
The endpoint inverse consequently tends to $+i\epsilon_0c\omega$.
Its anti-Hermitian inverse contribution, with the convention
$\Pi_\infty^r=\tfrac12[(v^a)^{-1}-(v^r)^{-1}]$, is
$-i\epsilon_0c\omega$.
The $O(h^{-1})$ real term in the raw embedding $-K\lambda$
is canceled by the $-K$ endpoint term; dropping either one
would make the continuum limit meaningless.
The construction assumes the stated site-weighted inner product;
a different discretization must carry its own boundary weights.



\section{Lesser-field covariance and planar stress with several sources}\label{app:stress-sources}
We distinguish the result for a prescribed incoming wave from the
full nonequilibrium covariance of an emitting sheet.  Fix a frequency
$\omega>0$, transverse momentum $\mathbf q_\perp$, and one physical
polarization $\sigma$.  Let $a_L$ denote the incoming wave from the
left and $b_R$ the incoming wave from the right.  Let $b_L$ and $a_R$
be the corresponding outgoing amplitudes.  A stationary linear sheet
can be represented as
\begin{equation}
 \begin{pmatrix}b_L\\a_R\end{pmatrix}
 =\underbrace{\begin{pmatrix}r_L&t_R\\t_L&r_R\end{pmatrix}}_{\mathsf S_\sigma}
 \begin{pmatrix}a_L\\b_R\end{pmatrix}
 +\begin{pmatrix}\xi_L\\\xi_R\end{pmatrix}.
\end{equation}
The last vector represents radiation generated by fluctuating sheet
currents.  It vanishes for a lossless sheet in the simple passive
scattering model, but not for a finite-temperature absorbing sheet.
For mutually uncorrelated external reservoirs and sheet sources, put
\begin{equation}
 C_{\rm in}=\begin{pmatrix}n_L&0\\0&n_R\end{pmatrix},\qquad
 C_\xi=\langle\boldsymbol\xi\boldsymbol\xi^\dagger\rangle,
\end{equation}
where $n_L,n_R$ stand for the appropriately normalized positive-frequency
field covariances.  Then
\begin{equation}\label{eq:outcov}
 C_{\rm out}=\mathsf S_\sigma C_{\rm in}\mathsf S_\sigma^\dagger+C_\xi.
\end{equation}
This is the mode-space version of dressing independent source
correlations with retarded and advanced Green's functions.  In NEGF,
$C_\xi$ comes from $D^r\Pi_{\rm sheet}^<D^a$, while the incoming
covariances come from the two bath self-energies.  The proportionality
between $n_\lambda$ and the Bose function depends on the continuum
mode normalization, so it is important not to identify $n_\lambda$
with a photon occupation without that measure.

For the phasor normalization used in Sec.~\ref{sec:lesser-pressure},
the time-averaged normal stress in each vacuum region is
\begin{equation}
 \overline T_{zz}=-W_\sigma\bigl(\langle|a|^2\rangle+
 \langle|b|^2\rangle\bigr),\quad
 W_\sigma=\frac{\epsilon_0}{2}\frac{k^2}{k_0^2}.
\end{equation}
The net force on the sheet is the right-side stress minus the
left-side stress.  In terms of the two covariance matrices,
\begin{align}
 \frac{F_z}{\mathcal A}
 &=W_\sigma\left[n_L+(C_{\rm out})_{LL}
 -n_R-(C_{\rm out})_{RR}\right].\label{eq:force-cov}
\end{align}
Equation~(\ref{eq:force-cov}) includes radiation from both baths and
from the sheet, once the source covariance is specified.  For
one-sided illumination $n_R=C_\xi=0$, it reduces to
$W_\sigma n_L(1+|r_L|^2-|t_L|^2)$, reproducing the main-text
scattering result.  The full force is obtained by summing over
polarizations and integrating with $d\omega/(2\pi)$ and
$d^2q_\perp/(2\pi)^2$, using a consistent continuum mode
normalization.  This extension is a source-bookkeeping identity,
not a microscopic evaluation of $C_\xi$ for a particular material.
For equilibrium Casimir stresses, vacuum contributions and a
reference subtraction must be retained; the normally ordered
thermal formula alone is insufficient.



\section{Units, frequency symmetry, and limiting checks}\label{app:checks}
The free vector-potential Green's function is normalized through
$L_0v=\hat 1$, where $L_0=\epsilon_0\omega^2I-\mu_0^{-1}\nabla\times\nabla\times$.
In the one-dimensional reduction, $v(z,z')$ is a kernel per
longitudinal length and its sampled matrix inverse is not the same
object as the inverse of a continuum differential operator.
For a planar Fourier transform, the boundary bath is a kernel
\emph{per unit area}; in a spherical representation it is integrated
with the surface measure $dS=R^2d\Omega$.
These measures must be kept when comparing a point-sampled matrix
with a continuum Green's function.

\subsection{Explicit SI dimensional bookkeeping}
It is useful to distinguish the dimensions of a continuum kernel from
those of a sampled or partially Fourier-transformed matrix.  Let
$\mathrm{H}$ denote the SI unit henry (not the magnetic field).  Since
$\epsilon_0c^2=\mu_0^{-1}$, the following quantities have dimensions
\begin{align}
 [\epsilon_0\omega^2]&=\mathrm{H}^{-1}\mathrm{m}^{-1},
 & [\sigma\omega]&=\mathrm{H}^{-1}\mathrm{m}^{-1},\\
 [\epsilon_0c\omega]&=\mathrm{H}^{-1},
 & [\epsilon_0c^2k]&=\mathrm{H}^{-1},\\
 [\eta]&=\mathrm{s}^{-1},
 & [k]=[q_\perp]&=\mathrm{m}^{-1}.
\end{align}
The three-dimensional free kernel behaves as $v^r\sim\mu_0/r$,
so $[v^r(\mathbf r,\mathbf r')]=\mathrm{H}\,\mathrm{m}^{-2}$.
After Fourier transformation in the transverse coordinates, the
planar kernel has dimensions $[v^r(\mathbf q_\perp;z,z')]=\mathrm{H}$.
The sampled one-dimensional matrix has the same units.  Accordingly,
its matrix inverse and the boundary self-energies in the present
normalization have dimensions $\mathrm{H}^{-1}$.

The longitudinal contact term is consistent with this counting:
$[\delta(z-z')/(\epsilon_0\omega^2)]=\mathrm{H}$.
Likewise, the site-weighted discrete coupling
$K=\epsilon_0c^2/h$ has dimensions $\mathrm{H}^{-1}$,
whereas the scalar-lattice parameter $\kappa=c^2/a^2$ in
Ref.~\cite{PengYapZhangWang2017} has dimensions $\mathrm{s}^{-2}$.
They should not be compared without accounting for field normalization.

Finally, the Planck integrand
$\hbar\omega^3N(\omega)d\omega/(\pi^2c^3)$ has dimensions
$\mathrm{J}\,\mathrm{m}^{-3}$, and
$[J_{\mathrm{in}}/c]=[T_{zz}]=\mathrm{N}\,\mathrm{m}^{-2}$.
The dimensionless quantities used in the attenuation and angular
integrals include $\sigma/(\epsilon_0\omega)$, $\eta r/c$,
$q_\perp/k_0$, and $k/k_0$.

The following checks are useful for detecting convention errors:
\begin{align}
 \Pi_{\infty,1D}^r&=-i\epsilon_0c\omega &&(\omega>0),\\
 \Pi_{\infty,s}^r&=-i\epsilon_0c^2k,\\
 \Pi_{\infty,p}^r&=-i\epsilon_0c^2\frac{k_0^2}{k},\\
 \Pi_{\infty,S_R}^r&=-i\epsilon_0c\omega P_R.
\end{align}
The planar expression reduces to the one-dimensional coefficient
at normal incidence $q_\perp=0$, because $k=k_0=\omega/c$.
At fixed $q_\perp<k_0$, both planar eigenvalues have negative
imaginary parts for $\omega>0$, so
$\Gamma_\infty=i(\Pi_\infty^r-\Pi_\infty^a)$ is positive
semidefinite.  For $q_\perp>k_0$, the anti-Hermitian radiative
operator vanishes, but this does not mean that the full finite-distance
outgoing embedding operator vanishes.  At grazing incidence,
$k\to0$, the TM eigenvalue is singular; integrated observables
must be analyzed before taking that limit.

With the Fourier convention $e^{-i\omega t}$, reality of the
real-time response requires
\begin{equation}
 \Pi_\infty^r(-\omega)=\bigl[\Pi_\infty^r(\omega)\bigr]^*
\end{equation}
for real-space kernels.  In particular, extending the positive-frequency
planar expression to negative frequency requires the retarded branch
of $k(\omega)$, not the positive square root for every real $\omega$.
The full-frequency Bose function satisfies
$N(-\omega)=-[1+N(\omega)]$; this identity must be retained in
frequency-mixing calculations.


\section{Comparison with the 2017 scalar-photon surface bath}\label{app:peng-comparison}
The earlier scalar-photon model of Peng \emph{et al.} \cite{PengYapZhangWang2017} uses a negative-definite field Hamiltonian and defines a discrete Green's function with the sign convention [their Eq.~(B9)]
\begin{equation}
 [ (\omega+i0^+)^2 I-K_e] \check d^{\,r}=-I.
\end{equation}
Here $K_e$ is the usual tridiagonal positive lattice dynamical matrix with diagonal $2\kappa$ and off-diagonal $-\kappa$, where $\kappa=c^2/a^2$. This extra minus sign is a consequence of the scalar-field commutator convention in that paper. With $\lambda+\lambda^{-1}=2-(\omega+i0^+)^2/\kappa$, its retarded surface Green's function satisfies
\begin{equation}
 \check d_{00}^{\,r}=\frac{\lambda_<}{\kappa},\qquad
 \check\Pi_L^r=\kappa\check d_{00}^{\,r}\kappa=\kappa\lambda_<,
\end{equation}
corresponding to their Eqs.~(B13)--(B17). The lesser self-energy is obtained with the scalar-field thermal convention and the normalization $\phi_j=u_j/\sqrt{s a}$, $s=\epsilon_0\mathcal A/c^2$; thus $\Pi_L^<=\lim_{a\to0}s a\check\Pi_L^<$. Their result is expressed in terms of the scalar occupation and the factor $\Omega$ [their Eq.~(B18)].

Our transverse-photon convention instead uses a site-weighted inverse with diagonal $\epsilon_0h\omega^2-2K$ and off-diagonal $K=\epsilon_0c^2/h$. Accordingly, the surface Green's function in Appendix~\ref{app:chain-detail} obeys $g_s=-\lambda/K$, and the embedding term is $\Sigma_B=-K\lambda$. The two sign conventions are therefore not in contradiction. In the transverse calculation, the combination of the embedding term and the central endpoint diagonal yields
\begin{equation}
 \epsilon_0h\omega^2-K-\Sigma_B
 =\epsilon_0h\omega^2+K(\lambda-1)
 \longrightarrow i\epsilon_0c\omega,
\end{equation}
for positive frequency and the outgoing root. Extracting its anti-Hermitian inverse contribution gives the radiative self-energy $\Pi_\infty^r=-i\epsilon_0c\omega$ per endpoint.

This comparison identifies the shared method---discretization, retarded surface Green's function, and continuum limit---without equating the physical scalar-potential reservoir of Ref.~\cite{PengYapZhangWang2017} with the transverse radiation bath considered here. The comparison also illustrates why a bare discrete self-energy, its sign convention, and its continuum boundary normalization must be stated separately.


\section{Historical development and source notes}\label{app:history}
The present paper draws on the author's handwritten research notes developed between 2016 and 2024. Notes dated February 2016 investigate the discretization of the one-dimensional photon field into a harmonic chain coupled to semi-infinite baths, as well as the bath self-energies and their continuum limits. Notes dated March 2019 discuss the bath at infinity in connection with photon NEGF conservation laws and radiation through a large spherical boundary. Subsequent notes from 2023 and 2024 investigate the planar geometry, the inverse Green's function method using $V_-$, $V_0$, and $V_+$, the longitudinal contact term, and the surface Green's function approach. These notes contain exploratory calculations and unresolved intermediate steps; the derivations presented here have been reorganized and checked.

Related published work includes the scalar-photon surface Green's function construction of Peng \emph{et al.} (2017) \cite{PengYapZhangWang2017}, the spherical bath formulation of Zhang \emph{et al.} (2022) \cite{ZhangZhuZhangWang2022}, the planar Floquet-radiation treatment of Zhu \emph{et al.} (2026) \cite{ZhuRenPanTangZhangWang2026}, and the systematic bath-at-infinity treatment of Tang, Ren, and Wang (2026) \cite{TangRenWang2026}. The material was also presented at the PTES/WTT meeting in Harbin in July 2026. The historical notes are mentioned here to document the development of the present exposition, without making claims of priority over earlier literature.


\end{document}